\documentclass[12pt,epsf]{article}

\usepackage{graphicx}
\usepackage{amsfonts}
\usepackage[usenames]{color}
\usepackage{amsmath}
\usepackage{amssymb}

\def\Z{\mathbb{Z}}
\def\R{\mathbb{R}}
\def\C{\mathbb{C}}

\def\Re{\mathrm{Re}}

\begin{document}
\baselineskip 0.6cm
\newcommand{\gsim}{ \mathop{}_{\textstyle \sim}^{\textstyle >} }
\newcommand{\lsim}{ \mathop{}_{\textstyle \sim}^{\textstyle <} }
\newcommand{\vev}[1]{ \left\langle {#1} \right\rangle }
\newcommand{\bra}[1]{ \langle {#1} | }
\newcommand{\ket}[1]{ | {#1} \rangle }
\newcommand{\Dsl}{\mbox{\ooalign{\hfil/\hfil\crcr$D$}}}
\newcommand{\nequiv}{\mbox{\ooalign{\hfil/\hfil\crcr$\equiv$}}}
\newcommand{\nsupset}{\mbox{\ooalign{\hfil/\hfil\crcr$\supset$}}}
\newcommand{\nni}{\mbox{\ooalign{\hfil/\hfil\crcr$\ni$}}}
\newcommand{\EV}{ {\rm eV} }
\newcommand{\KEV}{ {\rm keV} }
\newcommand{\MEV}{ {\rm MeV} }
\newcommand{\GEV}{ {\rm GeV} }
\newcommand{\TEV}{ {\rm TeV} }

\def\diag{\mathop{\rm diag}\nolimits}
\def\tr{\mathop{\rm tr}}
\def\Tr{\mathop{\rm Tr}}

\def\Spin{\mathop{\rm Spin}}
\def\SO{\mathop{\rm SO}}
\def\O{\mathop{\rm O}}
\def\SU{\mathop{\rm SU}}
\def\U{\mathop{\rm U}}
\def\Sp{\mathop{\rm Sp}}
\def\SL{\mathop{\rm SL}}
\def\simgt{\mathrel{\lower2.5pt\vbox{\lineskip=0pt\baselineskip=0pt
           \hbox{$>$}\hbox{$\sim$}}}}
\def\simlt{\mathrel{\lower2.5pt\vbox{\lineskip=0pt\baselineskip=0pt
           \hbox{$<$}\hbox{$\sim$}}}}

\def\change#1#2{{\color{blue} #1}{\color{red} [#2]}\color{black}\hbox{}}


\begin{titlepage}

\begin{flushright}
LTH 793 \\
CALT-68-2690 \\
UT-08-18 \\
IPMU 08-0035
\end{flushright}

\vskip 0.8cm
\begin{center}
 {\large \bf GUT Relations from String Theory Compactifications} 

\vskip 1.0cm

${}^a$Radu Tatar and ${}^{b,c}$Taizan Watari

\vskip 0.8cm
${}^a$ {\it Division of Theoretical Physics, Department of Mathematical
 Sciences, The University of Liverpool, Liverpool, L69 3BX, England,
 U.K} \\

\vskip 0.1cm
${}^b$ {\it Department of Physics, University of Tokyo, Tokyo, 
    113-0033, Japan} \\
\vskip 0.1cm

${}^c$ {\it Institute for the Physics and Mathematics of the Universe 
   (IPMU), University of Tokyo, Kashiwa-no-ha 5-1-5, 277-8592, Japan} \\

\vskip 1cm

\abstract{Wilson line on a non-simply connected manifold is 
a nice way to break SU(5) unified symmetry, and to solve 
the doublet--triplet splitting problem. 
This mechanism also requires, however, that the two Higgs doublets are 
strictly vector-like under all underlying gauge symmetries, and 
consequently there is a limit in a class of modes 
and their phenomenology for which the Wilson line can be used. 
An alternative is to turn on a non-flat line bundle 
in the $\U(1)_Y$ direction on an internal manifold, which does not 
have to be non-simply connected.
The $\U(1)_Y$ gauge field has to remain in the massless spectrum, and 
its coupling has to satisfy the GUT relation. In string theory 
compactifications, however, it is not that easy to satisfy these 
conditions in a natural way; we call it $\U(1)_Y$ problem. 
In this article, we explain how the problem is solved in some parts 
of moduli space of string theory compactifications. 
Two major ingredients are an extra strongly coupled U(1) gauge field 
and parametrically large volume for compactification, which is also 
essential in accounting for the hierarchy between the Planck scale 
and the GUT scale. Heterotic-M theory vacua and F-theory vacua 
are discussed. 
This article also shows that the toroidal orbifold GUT approach 
using discrete Wilson lines corresponds to the non-flat line-bundle 
breaking above when orbifold singularities are blown up.
Thus, the orbifold GUT approach also suffers from the $\U(1)_Y$ problem, 
and this article shows how to fix it. } 

\end{center}
\end{titlepage}


\section{Introduction}

The gauge coupling unification of the minimal supersymmetric 
standard model (MSSM) is the biggest (phenomenological) motivation 
to study supersymmetric unified theories. The SU(5)$_{\rm GUT}$ 
unified symmetry is broken down to the standard-model gauge 
group $\SU(3)_C \times \SU(2)_L \times \U(1)_Y$ 
without reducing the rank of the gauge group, 
when an expectation value is turned on for a scalar field 
in the SU(5)$_{\rm GUT}$ adjoint representation.

For higher-dimensional supersymmetric theories such as
geometric compactification of the superstring theory, 
there always exists $\SU(5)_{\rm GUT}$ gauge field 
with polarization pointing to the directions of internal manifold, 
and a Wilson line in the U(1)$_Y$ direction can play the role 
of the D = 4 scalar field in the adjoint representation.
The Wilson lines can be introduced only in a manifold $Z$ 
with a non-trivial homotopy group $\pi_1(Z) \neq \{ 1 \}$ 
\cite{Hosotani, CHSW, WittenWilson}. 
The Wilson lines in the U(1)$_Y$ direction, or equivalently
the flat bundles, break the SU(5)$_{\rm GUT}$ symmetry, 
get rid of gauge bosons in the off-diagonal blocks from the massless 
spectrum and allow the spectrum of coloured Higgs multiplets to be 
different from that of Higgs doublets. 

Since those goals can be achieved also by line bundles that are 
not flat, one could think of compactification on a simply connected 
manifold with a line bundle turned on in the U(1)$_Y$ direction, instead.
Many models fall into this category, including toroidal orbifold 
compactification 
\cite{Het-orb-A, Het-orb-B, Het-orb-C, Het-orb-D, Het-orb-E}\footnote{
Models leaving $\SU(3)_C \times \SU(2)_L$ (and some $\U(1)$'s) symmetry 
have been discussed in free-fermion formalism as well \cite{FF}. 
Relation between the free-fermion formalism and orbifold
compactification is studied in \cite{FF-orbifold}.} and 
$\SU(5) \times \U(1)_Y$ bundle compactification 
of Heterotic $E_8 \times E'_8$ string theory \cite{WittenSU(3),Munich0603} and 
Calabi--Yau orientifold compactification models of Type IIB 
string theory \cite{IIB-local-Madrid, Berenstein, 
IIB-local-Verlinde, Wijnholt}.

The problem of this approach with non-flat line bundles is that 
U(1)$_Y$ gauge field in the SU(5)$_{\rm GUT}$ symmetry 
(and hence U(1)$_{\rm QED}$) generically does not remain massless. 
This problem can be avoided by starting from a gauge group larger 
than SU(5)$_{\rm GUT}$, such as U(6) in a model
of Type IIB compactification \cite{IIB-local-Madrid}, or 
$E_8 \times E_8$ in Heterotic compactification \cite{Munich0603}. 
The massless U(1)$_Y$ gauge field below 
the Kaluza--Klein scale is a linear combination of the 
ordinary U(1)$_Y$ gauge field in the SU(5)$_{\rm GUT}$ gauge 
group and an additional U(1) symmetry contained in the larger 
gauge group. The gauge coupling constant of the low-energy 
U(1)$_Y$ gauge field is, however, weakened due to the mixture
of the additional U(1) gauge field, and the successful prediction 
of the gauge coupling unification is lost. 
The primary goal of this note is to show that the gauge coupling 
unification is restored in certain region (limit) of moduli space.

We are not only trying to explore just another class 
of string vacua with successful gauge coupling unification. Note that 
Wilson lines can be a solution to the doublet--triplet splitting 
problem only when a pair of Higgs doublets $H_u$ and $H_d$ is 
completely vector like under the underlying gauge symmetry 
such as $E_8$ (and in fact, $E_8$ is the only candidate of the underlying 
gauge symmetry if we assume $\SU(5)_{\rm GUT}$ unification and 
the vector-like nature of $H_u$ and $H_d$; see \cite{TW1}). 
In the Heterotic $E_8 \times E_8$ string theory, 
for instance, the Higgs multiplets $H({\bf 5})$ and $\bar{H}(\bar{\bf 5})$ 
may originate from 
$H^1(Z;\wedge^2 \overline{V_5})\simeq H^2(Z;\wedge^2 V_5^\times )$ and 
$H^1(Z; \wedge^2 V_5)$, respectively, where $Z$ is a Calabi--Yau 3-fold, 
$V_5$ is a rank-5 vector bundle in one of $E_8$ and 
$\overline{V_5}=V_5^\times$ its dual bundle. 
In this case, $H({\bf 5}) \supset H_u$ and 
$\bar{H}(\bar{\bf 5}) \supset H_d$ are vector-like not only under 
$\SU(5)_{\rm GUT}$ but also under the structure group $\SU(5)$. 
A flat bundle ${\cal L}_Y$ 
can be turned on in the U(1)$_Y$ direction, when $(Z,V_5)$ has 
an isometry group $\Gamma$ that acts freely on $Z$. 
The index theorem says that  
\begin{eqnarray}
 \# H_u - \# H_d & = & \chi(Z/\Gamma ; \wedge^2 V_5 \otimes 
   {\cal L}_Y^{-1/2})
    = \frac{1}{\# \Gamma} \chi(Z;\wedge^2 V_5), \\
 \# H_c({\bf 3}) - \# \bar{H}_c(\bar{\bf 3}) & = & 
\chi(Z/\Gamma ; \wedge^2 V_5 \otimes 
   {\cal L}_Y^{+1/3})
    = \frac{1}{\# \Gamma} \chi(Z;\wedge^2 V_5),
\end{eqnarray}
and hence coloured Higgs multiplets can be absent in the low-energy  
spectrum (that is, $\# H_c=0$ and $\# \bar{H}_c = 0$), 
while we have a pair of massless Higgs doublets, $\# H_u = \# H_d = 1$. 

If $H_u$ and $H_d$ originate from bundles that are not dual, 
on the other hand, the index theorem has to be applied separately for 
the bundle of $H_u$ and that of $H_d$. Suppose, say, that they are 
identified with cohomology groups 
$H^1(Z/\Gamma; U_{H_u} \otimes {\cal L}_Y^{+1/2})$ and 
$H^1(Z/\Gamma; U_{H_d} \otimes {\cal L}_Y^{-1/2})$ for some 
bundles $U_{H_u}$ and $U_{H_d}$ on $Z$, respectively.
Now, $H_u$ and $H_d$ are not vector-like under the structure groups 
of the vector bundles.
In this case, $\# H_u$ and $\# H_d$ are directly related to the Euler 
characteristics 
$- \chi(Z/\Gamma; U_{H_u} \otimes {\cal L}_Y^{+1/2})$ and 
$- \chi(Z/\Gamma; U_{H_d} \otimes {\cal L}_Y^{-1/2})$.
If the symmetry breaking of $\SU(5)_{\rm GUT}$ were due to a flat bundle 
${\cal L}_Y$ in the $\U(1)_Y$ direction, then the Euler characteristic 
of the bundle in the doublet parts and the triplets part cannot be
different, 
\begin{eqnarray}
\chi(Z/\Gamma; U_{H_u} \otimes {\cal L}_Y^{+1/2}) & = & 
\chi(Z/\Gamma; U_{H_u} \otimes {\cal L}_Y^{-1/3}) = 
\chi(Z; U_{H_u})/\# \Gamma, \\
\chi(Z/\Gamma; U_{H_d} \otimes {\cal L}_Y^{-1/2}) & = & 
\chi(Z/\Gamma; U_{H_d} \otimes {\cal L}_Y^{+1/3}) = 
\chi(Z; U_{H_d})/\# \gamma, 
\end{eqnarray}
because flat bundles do not contribute to the Euler characteristics.
Thus, if there is a pair of Higgs doublets $H_u$ and $H_d$ 
in low energy spectrum, and there is only a pair, then there is also 
a pair of Higgs triplets at low energies. Gauge coupling unification
is no longer expected in the presence of this additional triplets 
in ths spectrum.
If the $\SU(5)_{\rm GUT}$ symmetry is broken by a non-flat line bundle 
in the $\U(1)_Y$ direction, however, the chirality in the doublet 
part and the triplet part can be different, and there can be no triplets
at low energies; such compactifications are consistent with the gauge
coupling unification.
(hereafter, whenever we say a line bundle in this article, 
it is meant to be non-flat unless specifically mentioned 
as a flat bundle.) 

Models with non-vector-like two Higgs doublets has a natural 
mechanism to bring dimension-5 proton decay operators under 
control \cite{TW1,KNW}. 
A pair of Higgs multiplets being completely vector-like is the 
essence of the dimension-5 proton decay problem, and hence 
this problem is always an issue for the $\SU(5)_{\rm GUT}$ symmetry 
breaking using the Wilson line.
Although the dimension-5 operators 
can be eliminated by imposing an extra discrete symmetry for 
this special purpose, probability of finding such a symmetry
in a landscape of vacua is very small.\footnote{
There would hardly be an anthropic argument for such a discrete
symmetry, because it seems there is nothing wrong with a proton 
life time of order $10^{28}$ years. cf. \cite{Dine}.}
Thus, there exists a phenomenological motivation to study 
the $\SU(5)_{\rm GUT}$ symmetry breaking due to a line bundle 
in the $\U(1)_Y$ direction.

This article is organized as follows. 
Section~\ref{ssec:Y-problem} explains why it is difficult 
in wide class of string compactification to get a massless $\U(1)_Y$ 
gauge field while maintaining the gauge-coupling unification.
We see in section~\ref{ssec:Idea}, however, 
that this generic problem can be solved by assuming an extra 
strongly coupled U(1) gauge theory; the disparity between the
strongly coupled U(1) sector and the visible perturbative 
$\SU(5)_{\rm GUT}$ sector can be attributed to a parametrically 
large volume of compactification, which also accounts for 
the hierarchy between the unification scale and the Planck 
scale \cite{IWY,WY}.\footnote{An $E_r$-type underlying symmetry 
is essential in obtaining the Yukawa couplings as explained in \cite{TW1}, 
but not in the $\SU(5)_{\rm GUT}$ symmetry breaking. Thus,
although presentation of \cite{IWY,WY} uses Type IIB string theory, 
it does not mean that the idea cannot be extended to F-theory.}
This observation is elaborated in sections~\ref{sec:Het} 
and \ref{sec:F}, by using the compactifications of Heterotic string 
and F-theory, respectively. Along the way, we will also see that 
the idea of containing $\U(1)_Y$ flux in a local region in 
the internal space \cite{HN} is useful in bringing threshold 
corrections under control. Presentation of \cite{HN}
(and orbifold-GUT papers that followed) is based exclusively on 
toroidal orbifold compactification (of the Heterotic $E_8 \times E_8$ 
string theory), but we find a way to implement the idea 
in general string theory compactifications.

The appendix, which constitutes a big part of this paper, is somewhat 
independent from the main text of this article. It explains 
how the toroidal orbifold compactification 
is understood as certain limits of Calabi--Yau compactification. 
Heterotic orbifold-GUT approach in the last several years often make 
use of ``discrete Wilson lines'' in breaking the $\SU(5)_{\rm GUT}$ 
symmetry, and the primary purpose of the appendix is to clarify 
the meaning of discrete Wilson lines of toroidal orbifold
compactification in terms of Calabi--Yau compactification. 
The discrete Wilson lines in toroidal orbifolds are totally different 
from the Wilson lines associated with finite discrete homotopy group 
$\pi_1(Z)$ of non-simply connected Calabi--Yau $Z$. They should be 
understood as special cases (and special corners of moduli space) 
of non-flat line bundles in the $\U(1)_Y$ direction on Calabi-Yau
compactifications.

Thus, orbifold GUT models also suffer from 
the $\U(1)_Y$ problem in section~\ref{ssec:Y-problem}, and 
this problem is solved as we explain in this article. 
Because the idea of orbifold GUT has received attention for the last 
several years from much wider community, the appendix is 
pedagogically presented. The appendix~\ref{ssec:continuous} 
shows that the ``continuous Wilson lines'' in toroidal orbifold 
compactifications corresponds to vector-bundle mouli of smooth 
Calabi--Yau compactifications, and has nothing to do with Wilson 
lines associated with $\pi_1(Z) \sim \Z$.
\footnote{Although the contents 
of the appendix~\ref{ssec:continuous} is irrelevant to the main text, 
we include the contents of the appendix~\ref{ssec:continuous} 
in this note, because little effort beyond the 
appendix~\ref{ssec:discrete} is necessary, yet we expect that some 
people are interested in geometric interpretation of various aspects 
of toroidal orbifolds.}

As we were finishing this work, an article \cite{Vafa2} was 
posted on the web, which also discusses $\SU(5)_{\rm GUT}$ breaking 
due to a line bundle in the $\U(1)_Y$ direction.
There, an idea of \cite{IIB-local-Verlinde} in perturbative Type IIB 
string theory is generalized to F-theory compactifications, and 
explicit examples of geometry are given. Thus, a solution to the
$\U(1)_Y$ problem in this article (and in \cite{IWY,WY}) is 
different from those in \cite{IIB-local-Verlinde, Vafa2}.
We have also learnt that Donagi and Wijnholt have been working 
on a related subject (\cite{DW2}).

\section{The U(1)$_Y$ Problem and an Idea to Solve It}
\label{sec:2}

\subsection{The U(1)$_Y$ Problem}
\label{ssec:Y-problem}
\subsubsection{Massless U(1) Gauge Field}

Let us first consider the Heterotic $E_8 \times E_8$ theory compactified 
on a Calabi--Yau 3-fold $Z$ with vector bundles $V_5$ and $L_Y$
turned on in one of $E_8$. The structure group of $V_5$ is $\SU(5)_{\rm bdl}$, 
whose commutant in the $E_8$ symmetry is the SU(5)$_{\rm GUT}$ symmetry. 
The line bundle $L_Y$ is in the $\U(1)_Y \subset \SU(5)_{\rm GUT}$
direction. 
The $\SU(3)_C \times \SU(2)_L \times \U(1)_Y$ symmetry 
of the standard model is the commutant of 
the bundle structure group $\SU(5) \times \U(1)_Y$.
The gauge fields of the non-Abelian part of the unbroken symmetry, 
$\SU(3)_C \times \SU(2)_L$, remain massless below the Kaluza--Klein scale.

The U(1)$_Y$ gauge field, however, does not remain 
massless \cite{WittenSO(32), WittenSU(3), DSWW}.
The D = 10 action of the Heterotic string theory contains 
the kinetic term of the $B$-field  
\begin{equation}
 S = - \frac{1}{4\kappa^2} \int d^{10}x \sqrt{g_{10}} e^{-2 \phi} |H|^2; 
\qquad H = dB^{(2)} - \frac{\alpha'}{4} \left( 
  \tr_{E_8 \times E_8} \left(A F - \frac{2}{3}A A A \right) 
 - \omega_{\rm grav} \right),
\end{equation}
where $\omega_{\rm grav}$ is the Chern--Simons 3-form of gravity.
Fluctuations of the $B$-field of the form $b^k \omega_k$ are massless 
in the Kaluza--Klein reduction, where $b^k$ ($k = 1,\cdots,h^{1,1}$) are 
$D = 4$ scalar fields and $\omega_k$ form a basis of $H^{1,1}(Z)$ of a 
compact Calabi--Yau 3-fold $Z$. Their kinetic terms in the $D = 4$ 
effective theory are of the form\footnote{In addition to this 
generalized Green--Schwarz couplings of K\"{a}hler moduli 
chiral multiplets at tree level, there is also a 1-loop coupling 
for the dilation chiral multiplet \cite{DSW, ADS, DIS, DG(02)}.
Coefficients of the tree-level generalized Green--Schwarz couplings 
are worked out in \cite{Munich0504,Munich0603} in the Heterotic 
$E_8 \times E'_8$ string theory. (For SO(32) Heterotic string theory,
see \cite{MunichSO(32)}.)}
\begin{equation}
d^4 x \; {\cal L} =  d^4 x\; 
G_{kl} (\partial b^k - Q^k A)(\partial b^l - Q^l A); 
 \qquad c_1(L_Y) \propto \omega_k \, Q^k, 
\label{eq:B-A-Het}
\end{equation}
$G_{kl}$ is a metric on the K\"{a}hler moduli space \cite{Strominger,CO}, 
and $A$ is the U(1)$_Y$ gauge field.
Thus, a linear combination of these $B$-field fluctuations is absorbed 
to be the longitudinal mode of the U(1)$_Y$ gauge field. The kinetic
term  
above also contains the mass term of the U(1)$_Y$ gauge field.
Thus, whether the bundle $L_Y$ is flat ($c_1(L_Y) \propto \vev{dA} = 0$) 
or not leads to a big difference in phenomenology.

The same problem exists in Type IIB Calabi--Yau orientifold compactification.
Let us consider the Type IIB string theory compactified on a Calabi--Yau 
3-fold $X$ with a holomorphic involution ${\cal I}$; the Calabi--Yau 3-fold 
is modded by an orientifold projection associated with ${\cal I}$; 
$D7$-branes are wrapped on holomorphic 4-cycles, so that ${\cal N} = 1$ 
supersymmetry is preserved in $D = 4$ effective theory. If 5 $D7$-branes 
are wrapped on a holomorphic 4-cycle $\Sigma$ of $X$, the SU(5)$_{\rm GUT}$ 
gauge field propagates on $\Sigma$. Suppose that a line bundle $L_Y$ 
is turned on on $\Sigma$ in the U(1)$_Y$ direction in SU(5)$_{\rm GUT}$ 
symmetry. Then the SU(5)$_{\rm GUT}$ symmetry is broken to 
$\SU(3)_C \times \SU(2)_L \times \U(1)_Y$ symmetry of the standard model. 
Although the $\SU(3)_C \times \SU(2)_L$ part of the gauge field remains 
massless in this Type IIB compactification as well, the U(1)$_Y$ gauge
field does not. The Wess--Zumino action on $\Sigma$ contains 
\begin{equation}
S_{CS; \Sigma} = \int_{\R^{3,1} \times \Sigma} dC \tr e^{\frac{F}{2\pi}}  
 \propto  \int_{\R^{3,1}} A \wedge d \, c^{m} 
    \int_{\Sigma} \omega_m \wedge c_1(L_Y) + \cdots,
\label{eq:C-A-IIB}
\end{equation}
where $D = 4$ 2-form fields $c^m$ describe massless fluctuations of 
the Ramond--Ramond 4-form field $C^{(4)} \sim c^m \omega_m$. $A$ is 
the U(1)$_Y$ gauge field. Thus, a linear combination of the $D = 4$ 
Hodge dual of the 2-forms $c^m$ is absorbed to be the longitudinal mode 
of the U(1)$_Y$ gauge field. The U(1)$_Y$ gauge field becomes massive, 
and so does the QED gauge field. 
This is a problem in the context of large volume compactification, 
e.g. \cite{IIB-Madrid} in toroidal orbifolds and e.g. \cite{LouisIIB} 
in orientifolded Calabi--Yau 3-folds in general.

These phenomena in the Heterotic theory and Type IIB theory are 
related by the string duality.
It is the $B$-field fluctuation of the form 
$b^k \omega_k \propto Q^k \omega_k$ 
in the Heterotic theory that is mixed with the U(1)$_Y$ gauge field. 
Roughly speaking, it corresponds to a fluctuations of the Ramond--Ramond 
2-form field $C^{(2)} \sim \hat{c}^k \omega_k \propto Q^k \omega_k$ 
in the Type I string theory, where $\hat{c}^k$ are $D = 4$ scalar fields, 
and then to $C^{(4)} \sim c^k (*\omega_k) \propto Q^k (* \omega_k)$ in 
Type IIB string theory, where the Hodge dual $*$ is taken in a complex 
2-fold $B$ that the Heterotic and Type IIB string theory share in the duality.

The above argument, however, does not mean that it is impossible to 
obtain a massless U(1) gauge field in the low-energy spectrum.
Each line bundle in a compactification leaves a U(1) gauge field, 
and each massless fluctuation of the $B$-field or Ramond--Ramond field 
couples to a linear combination of those U(1) gauge field through 
the generalized Green--Schwarz mechanism \cite{DSW}.
If there is an abundant supply of U(1) gauge fields compared with 
the number of the bulk moduli fields, the U(1) gauge fields with no 
moduli-field counterpart remain massless.\footnote{
In Type IIB compactification 
on an orientifold of a Calabi--Yau 3-fold $X$, there are 
$h^{1,1}(X)$ chiral multiplets containing fluctuations of Ramond--Ramond 
fields. In Heterotic compactification on a Calabi--Yau 3-fold, there are 
$h^{1,1}(Z)$ K\"{a}hler moduli chiral multiplets and one dilaton chiral 
multiplet. Under the Heterotic--F-theory duality, an elliptic-fibred $Z$ 
on a base 2-fold $B$ is mapped to a K3-fibred Calabi-Yau 4-fold 
$X'$ on $B$. Heterotic compactification
has an F-theory dual only when line bundles are trivial in the elliptic 
fibre direction (if they had non-trivial first Chern classes in the fibre 
direction, vector bundles would not be stable in the small fibre
limit). Thus, the K\"{a}hler 
moduli multiplet associated with the size of the elliptic fibre does not 
participate in the generalized Green--Schwarz mechanism. 
So, $(h^{1,1}(Z)-1)=h^{1,1}(B)$ K\"{a}hler moduli chiral multiplets and 
the dilaton chiral multiplet can absorb massless U(1) gauge fields 
in the Heterotic compactification. On the other hand, the Type IIB
compactification has $h^{1,1}(X) = h^{1,1}(B)+1$ chiral multiplets 
containing fluctuations of the Ramond--Ramond 4-form or 2-form. Thus, 
the same number of massless gauge fields are absorbed in both descriptions; 
otherwise those two descriptions were not dual! 
} 

Reference \cite{Munich0603} considered an $\SU(5) \times \U(1)_Y  
\times \U(1)_2$-bundle compactification of the Heterotic $E_8 \times E_8'$ 
string theory. The $\SU(5) \times \U(1)_Y$ bundle is in one of $E_8$, and 
another line bundle has a structure group U(1)$_2$ in $E_8'$.
The first Chern classes of the two line bundles are chosen to be parallel 
in $H^{1,1}(Z)$, 
so that the gauge fields of both U(1)$_Y$ and U(1)$_2$ couple to the 
one and the same linear combination of the $B$-field fluctuations: 
$B^{(2)} \propto c_1(L_Y) \propto c_1(L_2)$. This $B$-field fluctuation 
absorbs only a linear combination of the two massless U(1) gauge fields, 
and the other combination remains massless. This gauge field, which 
is a linear combination of gauge fields in the visible $E_8$ and the hidden 
$E_8'$, can be identified with the massless hypercharge gauge field. 
The ratio of the hypercharges of the fields in the visible sector
is determined by the charges of the original 
$\U(1)_Y \subset \SU(5)_{\rm GUT}$ gauge field; 
hence the standard explanation of the hypercharge quantization in 
SU(5) unified theories---the original motivation of unified theories---is 
maintained. 

The $\C^3/\Z_3$ model in Type IIB string theory 
in \cite{IIB-local-Madrid} breaks an SU(6) symmetry 
by turning on a line bundle.\footnote{The fractional D3-branes 
at the $\C^3/\Z_3$ singularity are not just D7-branes wrapped 
on the vanishing 4-cycle isomorphic to $\C P^2$. One of the three 
fractional D3-branes at this singularity should be interpreted as 
a two anti-D7-branes wrapped on the vanishing cycle with a rank-2 vector 
bundle turned on \cite{fractD3onC3Z3}. Thus, this model does not immediately 
fit to the discussion so far that is based on large-volume compactification. 
However, we only discuss symmetry breaking pattern and counting of massless 
U(1) gauge fields, and in that context, the difference between anti-D7 branes 
and D7-branes does not make an essential difference. The same is true for 
other models such as those in \cite{Berenstein, IIB-local-Verlinde, Wijnholt}.} 
The SU(6) symmetry is broken down to $\SU(3) \times \SU(2)\times \U(1) 
\times \U(1)$, the non-Abelian part of which is identified with those 
of the standard model gauge group. The chiral multiplet that describes 
the blow-up of the $\C^3/\Z_3$ singularity (and hence the size of the 
$\C P^2$ cycle) absorbs a linear combination of the two U(1) gauge fields, 
and the other linear combination remains massless. This massless 
gauge field can be identified with that of the hypercharge. 
Models in \cite{Berenstein, IIB-local-Verlinde, Wijnholt} adopt
essentially the same strategy in maintaining a massless U(1) gauge 
field in the low-energy spectrum. One should keep in mind that 
how many massless U(1) gauge field remains massless is a global issue. 

\subsubsection{Normalization of the Hypercharges}

The overall normalization of hypercharges---not just the quantized 
ratio among them---is also an important prediction of supersymmetric 
unified theories. The SU(5)$_{\rm GUT}$ GUT's predict that 
\begin{equation}
\frac{1}{(5/3) \alpha_Y} = \frac{1}{\alpha_{\rm GUT}} = 
\frac{1}{\alpha_C} = \frac{1}{\alpha_L},
\label{eq:GUT-rel}
\end{equation}
which is called the GUT relation. The factor $(5/3)$ in the 
denominator comes from 
\begin{equation}
 {\bf q}_Y = \diag \left(-\frac{1}{3},-\frac{1}{3},-\frac{1}{3},
                         \frac{1}{2},\frac{1}{2} \right), \qquad 
 \tr({\bf q}_Y^2) = \frac{5}{3}.
\end{equation}
In this article we imply $\tr = T_R^{-1} \tr_R$ for any representations 
$R$ and, in particular, $\tr = 2 \tr_F$ for fundamental representations 
of SU($N$) symmetries, $\tr = \tr_{\rm vect.}$ for vector representations 
of SO($2N$) symmetries and $\tr = (1/30) \tr_{\rm adj.}$ for 
adjoint representations of $E_8$ and $\SO(32)$.

Now, when considering the idea of section 2.1.1 to maintain a massless 
U(1) gauge field at low energies, the low-energy U(1) gauge symmetry 
is not exactly the same as the U(1) hypercharge of SU(5) unified theories. 
Let us first pick up an example in the Heterotic string compactification 
that we mentioned above. The linear combination of U(1) gauge fields 
that becomes massive is
\begin{equation}
  \left( d b^k - \frac{1}{4\pi} \tr ({\bf q}_Y^2 ) Q^k_Y A_Y 
               - \frac{1}{4\pi} \tr ({\bf q}_2^2 ) Q^k_Y A_2 \right)^2,
\label{eq:gen-GS}
\end{equation}
where 
\begin{equation}
 \left(\frac{F}{2\pi}\right)_{L_Y} = {\bf q}_Y c_1(L_Y) 
 = {\bf q}_Y Q^k_Y \omega_k, \qquad   
 \left(\frac{F}{2\pi}\right)_{L_2} = {\bf q}_2 Q^k_Y \omega_k.
\label{eq:LYL2}
\end{equation}
The assumption that $c_1(L_Y) \propto c_1(L_2)$ in $H^{1,1}(Z)$ allows 
us to express the first Chern classes by using the same set of linear 
combination coefficients $Q^k_Y$. 
Gauge fields $A_Y$ and $A_2$ have kinetic terms 
\begin{equation}
{\cal L} = - \frac{\tr ({\bf q}_Y^2)}{16\pi \alpha} F_Y^2 
           - \frac{\tr ({\bf q}_2^2)}{16\pi \alpha'} F_2^2
\end{equation} 
in the effective Lagrangian in $D=4$, where $\alpha$ and $\alpha'$ are 
effective fine structure constants in the visible and hidden sectors, 
and hence the canonically normalized gauge fields $\widetilde{A_Y}$ 
and $\widetilde{A_2}$ are obtained from $A_Y$ and $A_2$ by rescaling
them by $\sqrt{4\pi \alpha/\tr ({\bf q}^2_Y)}$ and 
$\sqrt{4\pi \alpha'/\tr ({\bf q}^2_2)}$ , respectively. 
Thus, the canonically normalized massive vector field 
$A_{\rm massive}$ and its orthogonal complement $A_{\tilde{Y}}$ 
are given in terms of $\widetilde{A_Y}$ and $\widetilde{A_2}$ by 
\begin{equation}
 \left(\begin{array}{l} A_{\rm massive} \\ A_{\tilde{Y}} \end{array}\right) =
 \frac{1}{\sqrt{\alpha \tr ({\bf q}_Y^2) + 
                \alpha'\tr ({\bf q}_2^2)} } 
 \left( \begin{array}{cc} 
        \sqrt{\alpha \tr ({\bf q}_Y^2)} & 
        \sqrt{\alpha'\tr ({\bf q}_2^2)} \\ 
        \sqrt{\alpha'\tr ({\bf q}_2^2)} & 
       -\sqrt{\alpha \tr ({\bf q}_Y^2)} 
        \end{array} 
 \right)
\left( \begin{array}{l} \widetilde{A_Y} \\ \widetilde{A_2} \end{array}\right).
\end{equation}
It is $A_{\tilde{Y}}$ that remains massless in low-energy effective 
theory.
Fields in the visible sector are coupled to the massless gauge field 
$A_{\tilde{Y}}$ through the original hypercharge gauge field 
$\widetilde{A_Y}$:  
\begin{equation}
 \partial - i {\bf q}_Y \sqrt{\frac{4\pi \alpha}{\tr ({\bf q}_Y^2) }} 
 \widetilde{A_Y} \rightarrow 
 \partial - i {\bf q}_Y \sqrt{\frac{4\pi \alpha}{\tr ({\bf q}_Y^2) }} 
 \frac{\sqrt{\alpha' \tr ({\bf q}_2^2)} }
      {\sqrt{\alpha \tr ({\bf q}_Y^2) + 
                \alpha' \tr ({\bf q}_2^2)} 
      } A_{\tilde{Y}}.
\end{equation} 
Thus, the gauge coupling constant of this massless hypercharge gauge field 
is given by 
\begin{equation}
 \frac{1}{\tr ({\bf q}_Y^2 ) \alpha_{\tilde{Y}}} 
 = \frac{1}{\alpha} + \frac{1}{\alpha'}
   \frac{\tr ({\bf q}_Y^2 )}{\tr ( {\bf q}_2^2 )};
\label{eq:alpha1-modfd}
\end{equation}
The above discussion is essentially the same as 
calculating the QED coupling constant in the Weinberg--Salam model.
In the weakly coupled Heterotic $E_8 \times E_8'$ string theory, 
the gauge coupling constants of the visible and hidden sector 
$E_8$, namely, $\alpha = \alpha_{\rm GUT} = \alpha_{E_8}$ and 
$\alpha' = \alpha_{E_8'}$ are the same at the tree level, and hence 
the second term in (\ref{eq:alpha1-modfd}) makes the hypercharge 
coupling constant weaker by of order 100\% \cite{Munich0603}. 
The GUT relation (\ref{eq:GUT-rel}) is not satisfied at all.

Let us now take an example of \cite{IIB-local-Madrid} in Type IIB 
string local singularity. There, $\U(1)_Y$ massless gauge field 
comes essentially from a subgroup of U(6) generated by
\begin{equation}
 {\bf q}_{\U(6)} = \diag \left(-\frac{1}{3},-\frac{1}{3},-\frac{1}{3},
  -\frac{1}{2}, -\frac{1}{2}, -1 \right).
\label{eq:U6}
\end{equation}
When all the six fractional D3-branes are assumed to have the 
same gauge coupling constant, the massless gauge field has 
a coupling constant given by 
\begin{equation}
 \frac{1}{\tr ({\bf q}_Y^2) \alpha_{\U(6)}} 
  = \frac{\tr ({\bf q}_{\U(6)}^2)}{\tr ({\bf q}_Y^2) \alpha_{C, L}}
  = \frac{11/5}{\alpha_{C, L}}.
\end{equation}
This is much smaller than those of $\SU(3)_C \times \SU(2)_L$, 
and this is because of the extra\footnote{
It is still possible to maintain the GUT relation (\ref{eq:GUT-rel}) in 
Type IIB compactification, if we give up Georgi--Glashow SU(5) unification. 
For example, one can take ${\bf q}_Y = {\bf q}_{B-L}/2 - {\bf q}_R$, 
where ${\bf q}_{B-L}$ is a generator of 
$\U(1)_{B-L} \subset \SU(3)_C \times \U(1)_{B-L} \SU(4)_C$, and 
${\bf q}_R$ that of $\U(1)_R \subset \SU(2)_R$. One could imagine 
that the gauge coupling constants of all of 
$\SU(4)_C \times \SU(2)_L \times \SU(2)_R$ are the same, 
if $4+2+2$ D7-branes are wrapped on one and the same holomorphic
4-cycle with parametrically large volume. If this is the case, 
then the GUT relation is satisfied because 
$\tr (({\bf q}_{B-L}/2-{\bf q}_R)^2) = 5/3$. 
(This fact was exploited in a Type IIA model \cite{BlumenhagenPS}.)
In order to obtain appropriate spectrum, however, some of the 
7-branes forming $\SU(4)_C \times \SU(2)_L \times \SU(2)_R$ have 
to be anti-7-branes, because there are sum rules in the net chirality 
of various representation if they are all D7-branes \cite{TW1}. 
Thus, the volume of the 4-cycle has to be comparable to the string 
length, and there, values of $B$-fields integrated over various 
2-cycles also have significant contributions to gauge couplings of 
(subgroups of) $\SU(4)_C \times \SU(2)_L \times \SU(2)_R$. 
Thus, it is not obvious whether such quiver standard model 
in Type IIB string theory naturally predicts the GUT relation. 
(cf \cite{Wijnholt})} 
last entry of (\ref{eq:U6}).

In summary, when the SU(5)$_{\rm GUT}$ symmetry is broken by 
a line bundle in the U(1)$_Y$ direction, the U(1)$_Y$ gauge field 
tends to be massive by absorbing the K\"{a}hler moduli along the 
direction of the first Chern class of the line bundle. By considering 
compactification with multiple line bundles, however, it is possible 
to keep a massless U(1) gauge field, under which the ratio of 
the charges of the standard-model particles is that of the hypercharges. 
The overall normalization of the new hypercharges, or equivalently 
the gauge coupling constant of the new massless hypercharge gauge field, 
is different from the standard prediction of SU(5)$_{\rm GUT}$ unified 
theories. We call it the U(1)$_Y$ problem.

\subsection{Solving the $\U(1)_Y$ Problem with a Strongly Coupled U(1) 
Gauge Field}
\label{ssec:Idea}

Gauge coupling constants are functions of moduli fields in string theory, 
and hence the GUT relation may be satisfied somewhere in the moduli space.
Since we know that the first term in (\ref{eq:alpha1-modfd}) satisfies 
the GUT relation, it is clear that the GUT relation is satisfied 
approximately, if the contribution from the second term in 
(\ref{eq:alpha1-modfd}) is negligible compared with the first term.
In other words, as long as the extra U(1) gauge symmetry that mixes into 
the hypercharge is strongly coupled at the compactification scale, 
the effective gauge coupling constant of hypercharge at low-energy 
is not very much different from the ordinary prediction of 
SU(5)$_{\rm GUT}$ unified theories. There are such field-theory models 
in the literature (eg. \cite{Yanagida}).

\begin{figure}[t]
\begin{center}
\includegraphics[width=.75\linewidth]{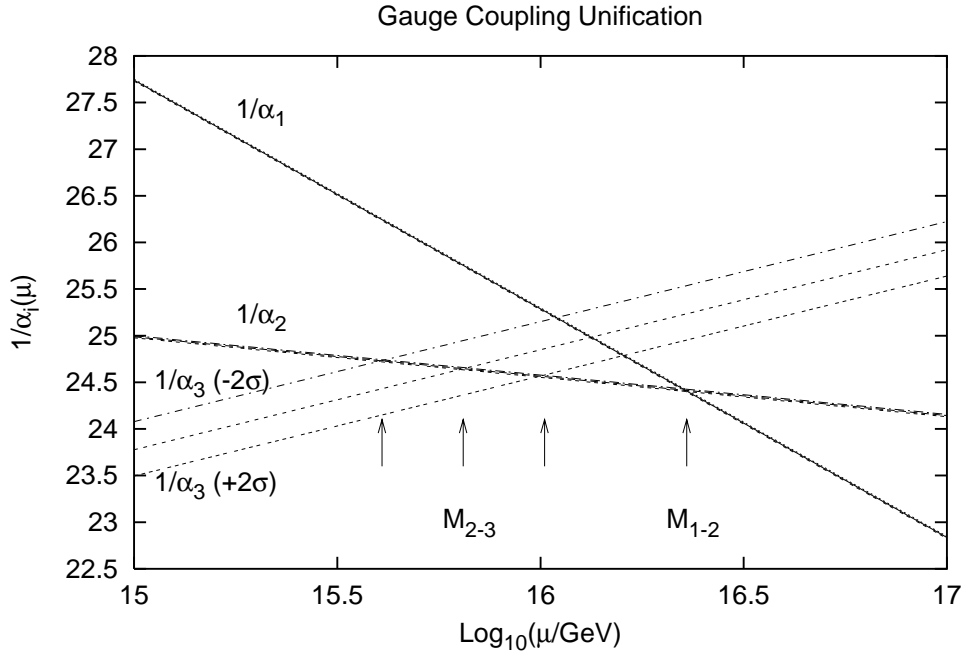} 
\end{center}
\caption{\label{fig:unif} This figure, borrowed from \cite{IW}, shows 
renormalization-group evolution of the three gauge coupling constants 
of the MSSM. Supersymmetry partners of the Standard-Model particles 
are assumed to be around 100 GeV--1 TeV, and 2-loop renormalization 
group equation was used for calculation. $\pm 2 \sigma$ error bar 
associated with the measurements of the QCD coupling is shown as 
the three parallel trajectories for $1/\alpha_3$. 
(See \cite{IW} for more details.)}
\end{figure}

As one can see in Figure \ref{fig:unif}, 
the three gauge coupling constants of the 
minimal supersymmetric standard model do not unify exactly at 
any energy scale around the GUT scale; at the energy scale 
$M_{2-3}$ in the figure, where $\alpha_C$ and $\alpha_L$ are equal, 
$(5/3) \times \alpha_Y$ is different from the others by 2--4\%.
Thus, the contribution from the second term in $(3/5)/\alpha_Y$ 
is phenomenologically acceptable. Furthermore, the extra contribution 
is supposed to be positive in $1/\alpha$, which is really the case
if the deviation from the GUT relation is due to the mixing with 
an extra strongly coupled U(1) gauge field.
We will see in the following sections that the extra U(1) is strongly coupled 
and hence the extra contribution to $1/\alpha_Y$ is small enough 
for some classes of string vacua in certain region of its moduli space.

Now one might wonder what is the point of maintaining the SU(5) 
unification. This is certainly a legitimate question. 
Unified theories can predict one of the three gauge coupling constants 
of $\SU(3)_C \times \SU(2)_L \times \U(1)_Y$ in terms of the other two, 
because there are only 2 parameters---the GUT scale and the unified 
gauge coupling constant. 
What is the point of considering a unified framework if one allows oneself 
to introduce an extra (moduli) parameter that change the U(1)$_Y$ gauge 
coupling? Predictability on the gauge coupling constants seems to be lost.
As we will see in the following sections, this is actually not the case. 
In the Heterotic--M-theoy compactification, the hidden sector gauge 
coupling is strong, due to the warping in the 11-th direction. 
In F-theory compactifications, which is motivated (as opposed to 
the perturbative Type IIB Calabi--Yau orientifold compactification) 
by the up-type Yukawa couplings \cite{TW1}, 
the dilaton vev cannot be small everywhere in the internal manifold.
Thus, having an extra strongly coupled U(1) gauge theory is extremely 
natural. Parametrically large volume for compactification is 
required in order to account for the little hierarchy between 
the GUT scale and the Planck scale, and a parametrically large 
volume to string length ratio can render the visible sector 
$\SU(5)_{\rm GUT}$ weakly coupled, in contrast to other strongly 
coupled sectors $\SU(5)_{\rm GUT}$ \cite{IWY, WY}.

From a perspective of phenomenology, the framework with a unified SU(5) 
and a strongly coupled extra U(1) symmetries says more than 
just having $\SU(3)_C \times \SU(2)_L \times \U(1)_Y$ massless 
gauge field at low energy with the GUT relation. 
The GUT gauge bosons exist around the 
energy scale of the gauge coupling unification, leading to 
dimension-6 proton decay. Since the rate of dimension-6 decay 
is proportional to the fourth power of the unification scale, 
the rate, and the proton lifetime is very sensitive to 
where the unification scale really is.
If we take a closer look at where the ``unification scale'' is, 
it is important to note that the extra contribution to $(3/5)/\alpha_Y$  
is always positive. 
Thus, ``the unification scale'' is more likely to be around $M_{2-3}$ 
in Figure~\ref{fig:unif} than $M_{1-2} \simeq 2 \times 10^{16} \, \GEV$ 
conventionally referred to as the GUT scale. 
Although one has to take account of threshold corrections 
and non-perturbative corrections in order to determine the GUT gauge boson 
mass (or the Kaluza--Klein scale) precisely, it is unlikely that the scale 
is as high as $M_{1-2}$ without an accidental cancellation between 
the threshold/non-perturbative corrections and the tree-level deviation 
from the GUT relation. This implies that the proton decay may be 
faster considerably than estimation based on $M_{1-2}$ as the GUT scale.
All the statements above on proton decay is valid whether the 
framework is implemented in the Heterotic--M-theory or in F-theory 
compactifications. See also related comments in the following sections.

\section{Heterotic-M Theory Vacua}
\label{sec:Het}

The Heterotic $E_8 \times E_8'$ string theory is compactified on 
a Calabi--Yau 3-fold $Z$ to yield a D = 4 effective theory 
with ${\cal N} = 1$ supersymmetry. 
Vector bundles $V_1$ and $V_2$ have to be turned on in both visible and 
hidden $E_8$ symmetries, so that 
\begin{equation}
  c_2(V_1) + c_2(V_2) = c_2 (TZ).
\label{eq:H-Bianchi}
\end{equation}
Apart from special cases, 
\begin{equation}
   \int_Z J \wedge \left(c_2(V_1) - \frac{1}{2}c_2(TZ) \right) 
 = - \int_Z J \wedge \left(c_2(V_2) - \frac{1}{2}c_2(TZ) \right)
\label{eq:asymmetry}
\end{equation}
does not vanish for a K\"{a}hler form $J$ of the Calabi--Yau 3-fold $Z$.
When (\ref{eq:asymmetry}) is not zero, it is known (as we review later) 
that the gauge coupling of one of the two $E_8$ gauge groups is stronger 
than that of the other $E_8$. 
For a large string coupling, $g_s$, the difference becomes significant, 
and in the limit of the largest possible $g_s$, 
one of the gauge couplings of D = 4 effective theory is really strongly 
coupled \cite{WittenStrongGs,CK1}.
Thus, if the $E_8$ gauge group with the weaker gauge coupling is identified 
the visible sector, $\alpha_{E_8} = \alpha_{\rm GUT}$, and the other $E_8'$ 
symmetry is strongly coupled,\footnote{
An unbroken subgroup of this $E_8$ symmetry may lead to dynamical 
supersymmetry breaking. The energy scale of the supersymmetry breaking 
$\Lambda_{DSB}$ is, however, determined by a combination 
$(2\pi/b_0 \alpha_{E_8'})$ where $b_0$ is the 1-loop beta function 
of the gauge coupling of the unbroken symmetry; 
the coupling $\alpha_{E_8'}$ alone does not determine the scale. 
Thus, the supersymmetry breaking scale can be much lower 
than the Kaluza--Klein scale when this hidden sector is nearly conformal, 
$b_0 \approx 0$. In model-building in F-theory, there is no such tight 
relation between the supersymmetry breaking scale and the deviation from 
the GUT relation. This may be regarded as a motivation for model building 
in F-theory.} and $1/\alpha_{E_8'}$ in (\ref{eq:alpha1-modfd}) is small; 
the GUT relation is maintained approximately. 
The purpose of this section is to check if this idea really works. 

\subsection{In Language of  the Weak Coupling Heterotic String Theory}

A vector bundle $V_5$ whose structure group is $\SU(5)_{\rm bdl} \subset E_8$ 
breaks the $E_8$ symmetry down to the commutant of the $\SU(5)_{\rm bdl}$, 
SU(5)$_{\rm GUT}$. The SU(5)$_{\rm GUT}$ symmetry is further broken 
down to $\SU(3)_C \times \SU(2)_L \times \U(1)_Y$ by turning on 
a line bundle $L_Y$ in the hypercharge direction. The $E_8$ super 
Yang--Mills fields of D = 10 Heterotic string theory yield all the 
gauge and matter multiplets except just one, U(1)$_Y$ vector multiplet.
The U(1)$_Y$ symmetry may remain unbroken as a global symmetry, but 
the gauge field absorbs a fluctuation of the $B$-field, and becomes 
massive. Whether the SU(5)$_{\rm GUT}$ symmetry is broken by a flat bundle 
or by a line bundle makes a big difference \cite{WittenSU(3)}.

References \cite{WittenSU(3), Munich0603} proposed a solution 
to this problem. Here, we briefly review the construction of
\cite{Munich0603} in order to set the notation in this article.

The (weakly coupled) $E_8 \times E_8'$ Heterotic string theory is 
compactified on a Calabi--Yau 3-fold $Z$, whose $\pi_1(Z)$ does not 
have to be non-trivial. A vector bundle $V_1$ is turned on 
in the visible sector $E_8$, which consists of a rank-5 vector bundle $V_5$ 
and a line bundle $L$. The D = 10 $E_8$ super Yang--Mills multiplet 
yields all the chiral multiplets necessary in supersymmetric standard model; 
see Table~\ref{tab:bndl-MSSM}.
\begin{table}[t]
\begin{center}
\begin{tabular}{r|ccccccc}
multiplets & $Q$ & $\overline{U}$ & $\overline{E}$ & $\overline{D}$ & $L$
  & $H_u$ & $H_d$ \\
\hline
bundles & $V_5 \otimes L^{\frac{1}{6}}$ & $V_5 \otimes L^{-\frac{2}{3}}$ & 
$V_5 \otimes L$ & $\wedge^2 V_5 \otimes L^{\frac{1}{3}}$ & 
$\wedge^2 V_5 \otimes L^{-\frac{1}{2}}$ & 
$\overline{\wedge^2 V_5} \otimes L^{\frac{1}{2}}$ & 
$\wedge^2 V_5 \otimes L^{-\frac{1}{2}}$
\end{tabular}
\caption{\label{tab:bndl-MSSM}Vector bundles of chiral multiplets in 
supersymmetric standard models. For a realistic model, the vector bundle 
$V_5$ cannot be generic; otherwise, there is a problem of dimension-4 
proton decay. For example, a $\Z_2$ symmetry (matter parity or R-parity) 
or an extension structure removes virtually all the dimension-4 proton 
decay operators \cite{TW1,TW2,KNW}. We do not go into details because 
such extra structures of the bundle $V_5$ are not essential to the 
gauge coupling unification, the main theme of this article. 
We used a notation $L$ for $L_Y$ in the text to save space in this table.} 
\end{center}
\end{table}
$\SU(3)_C \times \SU(2)_L$ gauge fields remain massless. 
A vector bundle $V_2$ in the ``hidden sector'' $E_8'$ 
should contain a line bundle $L_2$ (and possibly another bundle $V'$ 
whose structure group commutes with the U(1)$_2$ structure group of $L_2$) 
which satisfies 
\begin{equation}
c_1(L_2) \propto c_1(L_Y) \in H^{1,1}(Z).
\end{equation}
We set the normalization of the generator ${\bf q}_2$ for $L_2$ 
as in (\ref{eq:LYL2}), using $c_1(L_Y)$. 
The second Chern classes are given by 
\begin{eqnarray}
 c_2(V_1) & = & c_2(V_5) - \frac{\tr ({\bf q}_Y^2 )}{4} c_1(L_Y)^2, \\
 c_2(V_2) & = & c_2(V') - \frac{\tr ({\bf q}_2^2 )}{4} c_1(L_Y)^2, 
\end{eqnarray}
and they have to satisfy the consistency condition (\ref{eq:H-Bianchi}).
An explicit example of a Calabi--Yau 3-fold $Z$ and vector bundles on it 
is found in \cite{Munich0603}. 
In order to obtain the spectrum of supersymmetric standard model, 
bundles introduced so far have to satisfy 
\begin{equation}
\int_Z c_1(L_Y) \wedge c_2(TZ) = 0, \qquad \int_Z c_1(L_Y) \wedge c_2(V_5)= 0,
\qquad \int_Z c_1(L_Y)^3 = 0.
\end{equation}

Dimensional reduction of a Calabi--Yau compactification leaves a dilaton 
chiral multiplet $S$ and $h^{1,1}(Z)$ K\"{a}hler moduli chiral multiplets 
$T^k$ ($k=1,\cdots,h^{1,1}(Z)$):
\begin{eqnarray}
S & = & \frac{M_G^2 \alpha'}{4}\left(
 \frac{1}{e^{2\tilde{\phi}}}\frac{{\rm vol}(Z)}{\vev{{\rm vol}(Z)}} - i a \right), \\
T^k & = & \frac{1}{2\pi} \left( \alpha^k + i b^k \right)  \qquad 
    (k=1,\cdots,h^{1,1}(Z)), 
\end{eqnarray}
where $\tilde{\phi}$ and $a$ are dilaton fluctuation and
model-independent axion of the Heterotic string theory; 
$M_G \simeq 2.4 \times 10^{18} \, \GEV$ is given by  
\begin{equation}
 \frac{M_G^2}{2} = \frac{\vev{{\rm vol}(Z)}}{2 \kappa_{10}^2 g_s^2} 
 = \frac{\vev{{\rm vol}(Z)}}{(2\pi)^7 \alpha'^4 g_s^2}.
\end{equation}
$\alpha^k$ and $b^k$ parametrize the metric and $B$-field on $Z$ by 
\begin{equation}
 J = l_s^2 \alpha^k \omega_k, \qquad B = l_s^2 b^k \omega_k, 
\label{eq:JB}
\end{equation}
where $\omega_k$ ($k=1,\cdots,h^{1,1}(Z)$) are basis of $H^{1,1}(Z)$, 
and $J$ is a K\"{a}hler form\footnote{${\rm vol}(Z) = (1/3!) \int_Z J^3$ 
in this definition. Note that the K\"{a}hler form in \cite{WittenStrongGs} 
is $\omega = 
-i g_{\alpha \bar{\beta}} dz^\alpha \wedge d \bar{z}^{\bar{\beta}}$, 
different by a factor $-1$. } 
\begin{eqnarray}
 J & = & i g_{\alpha \bar{\beta}} dz^\alpha \wedge d \bar{z}^{\bar{\beta}}; \\
ds^2 & = & g_{\alpha \bar{\beta}} dz^\alpha \otimes d \bar{z}^{\bar{\beta}} 
     + g_{\alpha \bar{\beta}} d\bar{z}^{\bar{\beta}} \otimes d z^\alpha, 
\end{eqnarray}

The kinetic term of the $B$-field contains 
\begin{equation}
 \left| \left( d\, B^{(2)} - \frac{\alpha'}{4}
    \left( \omega_{{\rm YM} \, 1} + \omega_{{\rm YM} \, 2} \right)
 \right) \right|^2 \rightarrow 
 \left|\left(d \, b^k - \frac{Q^k_Y}{4\pi}\left(
   \tr ({\bf q}_Y^2 ) A_Y + \tr ({\bf q}_2^2) A_2 \right)
 \right) \omega_k \, l_s^2 \right|^2,
\end{equation}
where $A_Y$ and $A_2$ are gauge fields associated with the generators 
${\bf q}_Y$ and ${\bf q}_2$, respectively. 
A linear combination of vector multiplets, 
$V_{\rm massiv} \equiv \tr ({\bf q}_Y^2) V_Y + \tr ({\bf q}_2^2) V_2$, 
enters the K\"{a}hler potential as in 
\begin{equation} 
 K = 
 - M_G^2 \ln \left( \frac{1}{3!} \int_Z \tilde{J} \tilde{J} \tilde{J} \right),
   \qquad 
 \tilde{J} = -\pi l_s^2 \omega_k \left(T^k  + T^{k \, \dagger} 
   +\frac{Q^k_Y}{8\pi^2} V_{\rm massive}\right),
\end{equation}
and becomes massive. On the other hand, these vector multiplets do not 
have a similar coupling with the dilaton in the K\"{a}hler potential; 
although they could enter the K\"{a}hler potential as in
\begin{equation}
K = - \ln \left(S + S^\dagger + \frac{Q^S}{32\pi^2} V_{\rm massive} \right), \quad 
Q^S = \int_Z c_1(L_Y) \left( c_2(V_1) - \frac{1}{2} c_2(TZ) \right),
\end{equation}
$Q^S$ is proportional to U(1)$_1$--[non-Abelian]$^2$ mixed anomalies 
with SU(3)$_C$ (and SU(2)$_L$) as the non-Abelian gauge group (see also 
footnote \ref{fn:E8-SO32}), and hence 
vanishes in vacua with spectra of supersymmetric standard models.

Since only one linear combination, $V_{\rm massive}$ becomes massive, another 
linear combination of the gauge fields $A_Y$ and $A_2$ remains massless.
All the particles in Table~\ref{tab:bndl-MSSM} are charged under 
this massless U(1) gauge symmetry through its $A_Y$ component, and 
hence the ratio of the U(1) charges remains the same. This massless 
U(1)$_{\tilde{Y}}$ vector field is regarded as the hypercharge gauge
field of the Standard Model \cite{WittenSU(3), Munich0603}. 
The only problem of this solution is that the gauge coupling constants 
of $\SU(3)_C \times \SU(2)_L \times \U(1)_Y$---given as functions 
of moduli $S$ and $T^k$---do not satisfy (generically) 
the GUT relation. To see this, note that 
the gauge kinetic term of the two U(1) gauge fields $A_Y$ and $A_2$ is 
\begin{equation}
- \frac{1}{4} \left( \begin{array}{cc} F_Y & F_2 \end{array} \right) 
 {\rm Re}\; \left( \begin{array}{cc} 
  \tr ({\bf q}_Y^2) \left(S + T - \frac{\tr ({\bf q}_Y^2)}{3} A \right) &
  \frac{\tr ({\bf q}_Y^2) \, \tr ({\bf q}_2^2 ) }{6} A \\
  \frac{\tr ({\bf q}_Y^2) \, \tr ({\bf q}_2^2 ) }{6} A &   
  \tr ({\bf q}_2^2) \left(S - T - \frac{\tr ({\bf q}_2^2)}{3} A \right) 
                    \end{array} \right) 
             \left( \begin{array}{c} F_Y \\ F_2 \end{array} \right) 
\label{eq:gauge-kine-matrix}
\end{equation}
in the large volume limit, where 
\begin{eqnarray}
T & \equiv & \frac{1}{4} T^k \int_Z \omega_k \wedge 
    \left( c_2(V_1) - \frac{1}{2} c_2(TZ) \right), \label{eq:def-T}\\ 
A & \equiv & \frac{1}{4} T^k \int_Z \omega_k \wedge c_1(L_Y)^2.
\label{eq:def-A}
\end{eqnarray}
We only consider ${\rm Re} A \propto \int_Z J \wedge c_1(L_Y)^2 = 0$ 
for simplicity for the moment.\footnote{Later, we will see that it is an 
important assumption necessary for the gauge coupling unification.}
Following the process described in section 2, one can see that the 
massless linear combination is 
\begin{equation}
 A_{\tilde{Y}} \propto 
    \sqrt{ \frac{ \Re (S+T) }{ \Re (S-T) } } A_Y
  - \sqrt{ \frac{ \Re (S-T) }{ \Re (S+T) } } A_2, 
\end{equation}
and the gauge coupling constant is given by 
\begin{equation}
\frac{\frac{3}{5}}{g_{\tilde{Y}}^2} = {\rm Re}(S+T) + 
  \frac{\tr ({\bf q}_Y^2)}{\tr ({\bf q}_2^2 )} {\rm Re} (S-T).
\label{eq:asthesum}
\end{equation}
Note that $1/g^2_C$ and $1/g_L^2$ in the visible sector are given 
by 
\begin{equation}
 \frac{1}{g_C^2} = \frac{1}{g_L^2} = \Re f = \Re (S + T)
    \label{eq:f-nA-vis} 
\end{equation}
in the large volume limit.
When the hidden sector has an unbroken non-Abelian symmetry 
group, its gauge coupling constant 
is given by 
\begin{equation}
 \frac{1}{g^{'2}} = \Re f' = \Re \left( S + \frac{1}{4} T^k \int_Z \omega_k \wedge 
   \left( c_2(V_2) - \frac{1}{2} c_2(TZ) \right) \right)
    = \Re (S - T). 
\label{eq:f-nA-hid}
\end{equation} 
The U(1)$_{\tilde{Y}}$ gauge coupling in (\ref{eq:asthesum}) is given
just as the discussion in section~\ref{ssec:Y-problem}. 
In the weakly coupled Heterotic string theory, the tree-level coupling 
$\Re S$ dominates, with 1-loop corrections $\propto \Re T$ being subleading.
Thus, ignoring $\Re T$ in 
\begin{eqnarray}
 \frac{\frac{3}{5}}{g_{\tilde{Y}}^2} & = & \left(1 + 
  \frac{\tr ({\bf q}_Y^2) }{ \tr ({\bf q}_2^2 ) } \right) \Re S
 + \left(1 - 
  \frac{\tr ({\bf q}_Y^2) }{ \tr ({\bf q}_2^2 ) } \right) \Re T, \\
\label{eq:tree+1loop}
 & \simeq & \left(1 + 
  \frac{\tr ({\bf q}_Y^2) }{ \tr ({\bf q}_2^2 ) } \right) \frac{1}{g_{C,L}^2},
 \label{eq:}
\end{eqnarray}
the GUT relation is badly violated; the factor in the parenthesis on the 
right-hand side is different from 1 by of order unity for the model 
in \cite{Munich0603}.
If the 1-loop threshold correction, the second term in (\ref{eq:tree+1loop}),
were to partially cancel the tree level gauge coupling so that 
the gauge coupling constants of the MSSM apparently satisfy the 
GUT relation, it sounds very artificial.
This is the Heterotic-string version of the U(1)$_Y$ problem.

Reference \cite{WittenSU(3)} points out that the GUT relation is 
maintained approximately if $\tr ({\bf q}_2^2)$ is chosem much 
larger than $\tr ({\bf q}_Y^2)$. While this is true, we will see 
in the following, that the approximate GUT relation is actually 
maintained 
even if $\tr ({\bf q}_2^2)$ and $\tr ({\bf q}_Y^2)$ are comparable.

\subsection{In the Strongly Coupled Heterotic-M Theory}

\subsubsection{Strongly Coupled Hidden Sector}

When the gauge coupling constant in the hidden sector is way stronger 
than that of the visible sector for some reason, the second term of 
(\ref{eq:alpha1-modfd}) and (\ref{eq:asthesum}) is negligible, and the GUT 
relation is approximately satisfied; that was the idea of section 2, 
phrased in the context of the Heterotic string theory. 

Such a disparity between the gauge coupling constants naturally happen 
in strongly coupled Heterotic $E_8 \times E_8'$ string theory. 
The Bianchi identity of the NS--NS 2-form field requires that the 
total sum of the second Chern classes vanish, but they are not 
necessarily distributed equally to the visible and hidden sector. 
In general, (\ref{eq:asymmetry}) does not vanish,  
and the asymmetric distribution of the second Chern classes 
provide sources for the configuration of the Ramond--Ramond 3-form 
field in the bulk of the Heterotic-M theory.
\begin{equation}
 G_{\alpha \bar{\beta}\gamma \bar{\delta}}  = \left\{ 
 \begin{array}{rl}
\frac{(2\pi)^2}{\sqrt{2} \pi} \left(\frac{\kappa}{4\pi}\right)^{2/3}
\left(c_2(V_1) - \frac{1}{2} c_2(TZ) \right)_{\alpha \bar{\beta}\gamma \bar{\delta}} 
           &({\rm for~} 0 < x_{11} < \pi \rho), \\
- \frac{(2\pi)^2}{\sqrt{2} \pi} \left(\frac{\kappa}{4\pi}\right)^{2/3}
\left(c_2(V_1) - \frac{1}{2} c_2(TZ) \right)_{\alpha \bar{\beta}\gamma \bar{\delta}}
         &  ({\rm for~} -\pi \rho < x_{11} < 0). 
 \end{array}
\right.
\end{equation}
Coefficients are taken from \cite{Ovrut4}.
The non-zero 4-form field strength of the Ramond--Ramond field in the 
bulk, in turn, becomes the source of metric. 
The metric of $D = 11$ gravity is expanded as 
\begin{equation}
ds^2 = e^{b(x_{11})}dx^2 +
 2 (g_{\alpha \bar{\beta}} + h_{\alpha \bar{\beta}}) dz^\alpha d\bar{z}^{\bar{\beta}} +
 e^{k(x_{11})} dx_{11}^2,
\end{equation}
and at the linear order in $\kappa^{2/3}$, first order deformation
$b(x_{11})$ and $h(x_{11},z,\bar{z})$ follow the equations 
\begin{eqnarray}
\partial_{11} b & = & \frac{\sqrt{2}}{24} \alpha, \\
 \partial_{11} h_{\alpha \bar{\beta}} & = & 
- \frac{1}{\sqrt{2}} \left( i \Theta_{\alpha \bar{\beta}} - 
   \frac{1}{12}\alpha g_{\alpha \bar{\beta}} \right).
\end{eqnarray}
Here, $\Theta_{\alpha \bar{\beta}} := 2 i g^{\bar{\delta}\gamma} 
 G_{\alpha \bar{\beta}\gamma \bar{\delta}}$ and 
$\alpha = 2 i g^{\bar{\beta} \alpha} \Theta_{\alpha \bar{\beta}}$ 
as in \cite{WittenStrongGs}. 
If
\begin{equation}
\Theta_{\alpha \bar{\beta}} \propto g_{\alpha \bar{\beta}},
\label{eq:Theta}
\end{equation}
the last one above becomes $\partial_{11} h_{\alpha \bar{\beta}} 
= - \sqrt{2}/24 \alpha g_{\alpha \bar{\beta}}$, and hence 
the $(g_{\alpha \bar{\beta}} + h_{\alpha \bar{\beta}}$ part is 
of the form $e^f g_{\alpha \bar{\beta}}$ with $f$ satisfying 
$\partial_{11} f = - \sqrt{2}/24 \alpha$ \cite{CK1}. 
$k(z,\bar{z})$ should have the same $(z,\bar{z})$ dependence 
as $f$, and its $x_{11}$ can be chosen 
so that $k=f$ \cite{WittenStrongGs, CK1}.
Thus, the metric has the warped structure \cite{CK1}:
\begin{equation}
 ds^2 = e^{-f(x_{11})} e^{-\frac{2}{3}\phi} dx^2 + 
e^{f(x_{11})} (e^{-\frac{2}{3}\phi} g_{\alpha \bar{\beta}} dz^\alpha d\bar{z}^{\bar{\beta}} 
 + e^{\frac{4}{3}\phi} dx_{11}^2),
\end{equation}
where 
\begin{equation}
 e^{f(x_{11})} = \left(1 - \frac{\alpha}{8\sqrt{2}} x_{11}\right)^{\frac{2}{3}}.
\end{equation}
The volume of Calabi--Yau 3-fold varies over $x_{11}$, and 
in particular, decreases monotonically. It follows that 
\begin{equation}
 \frac{{\rm vol}(Z)|_{x_{11}=\pi \rho}}{{\rm vol} (Z)_{x_{11}=0}} = 
 \left(1 - \frac{\alpha}{8\sqrt{2}} \pi \rho \right)^2,
\end{equation}
and the gauge coupling constants of the visible and hidden 
sectors in $D=4$ effective theory are given by 
\begin{equation}
\frac{1}{\alpha_{\rm GUT}} = 
    \frac{{\rm vol}(Z)|_{x_{11}=0}}{(4\pi \kappa^2)^{\frac{2}{3}}}, 
\qquad 
\frac{1}{\alpha'_{\rm hidden}} = \frac{{\rm vol}(Z)_{x_{11}=\pi \rho}}
                         {(4\pi \kappa^2)^{\frac{2}{3}}} 
 = \frac{\left(1-\frac{\alpha}{8\sqrt{2}} \pi \rho\right)^2}{\alpha_{\rm GUT}}.
\end{equation}
Larger volume at $x_{11}=0$ makes the visible sector coupling 
weaker, while the hidden sector coupling remains strong 
\cite{WittenStrongGs,Stieberger,CK1}. 

The expression for the two gauge coupling constants 
in the weakly coupled Heterotic theory, (\ref{eq:f-nA-vis}) and 
(\ref{eq:f-nA-hid}) captures the warped factor effect. Indeed, 
\begin{eqnarray}
\frac{1}{g^2} - \frac{1}{g^{'2}} & = & 
\frac{1}{4\pi}\left(
  \frac{{\rm vol}(Z)|_{x_{11}=0}}{(4\pi \kappa^2)^{\frac{2}{3}}}
- \frac{{\rm vol}(Z)|_{x_{11}=\pi \rho}}{(4\pi \kappa^2)^{\frac{2}{3}}}
              \right) 
 \simeq \frac{1}{4\pi} 
     \frac{{\rm vol}(Z)|_{x_{11}=0}}{(4\pi \kappa^2)^{\frac{2}{3}}} 
     \frac{\pi \rho}{4\sqrt{2}} \alpha, \nonumber \\
 & = &  \frac{1}{4\pi}
     \frac{{\rm vol}(Z)|_{x_{11}=0}}{(4 \pi \kappa^2)^{\frac{2}{3}}}
     \frac{\pi \rho}{4\sqrt{2}} \frac{2 J \wedge G}{({\rm volume~form})}, 
   \nonumber \\
 & = & \frac{1}{4\pi} \frac{1}{(4\pi \kappa^2)^{\frac{2}{3}}}
    \frac{\pi \rho}{2 \sqrt{2}} \int_Z J \wedge G
   \propto  \frac{1}{4\pi l_s^2} \int_Z 
     J \wedge \left(c_2(V_1) - \frac{1}{2}c_2(TZ) \right)
\end{eqnarray}
in Heterotic-M theory language agrees with the result of weakly 
coupled Heterotic string theory, 
${\rm Re}(S+T) - {\rm Re}(S-T) = 2 {\rm Re} T$ 
(up to a proportionality factor).\footnote{They should agree without
a proportionality factor, but we have not succeeded in clarifying 
relation among various conventions in the literature.}
Here, higher order ${\cal O}(\kappa^{\frac{4}{3}})$ corrections 
are ignored.

Although the perturbative expansion of the Heterotic string 
theory is not reliable for $g_s > 1$, the gauge kinetic function 
is protected by holomorphicity. Only the tree and 1-loop level 
contributions exist, apart from non-perturbative corrections. 
They are given by $S\pm T$ at this level, and the holomorphicity 
of $f$ and $f'$ guarantees that their expressions are right 
as the perturbative part even in the strong coupling regime. 
It is true that the physical gauge coupling constants receive 
higher loop corrections despite the holomorphicity of ${\cal N}=1$ 
supersymmetry. However, such corrections arise only through 
the rescaling of the vector supermultiplets ($U(1)_{\tilde{Y}}$ and 
$\SU(3)_C \times \SU(2)_L$ in this case)
and super-Weyl transformation in rewriting Lagrangian in the 
Einstein frame.  The former only involve $\ln(g_{\tilde{Y}})$ and 
$\ln (g_C) = \ln (g_L)$ and are always small, 
while the latter is universal to all the gauge coupling constants. 
Thus, these corrections, which correspond to higher loops, 
are not the concern for us.

It appeared in language of weakly coupled Heterotic string theory 
that a fine-tuning between the tree-level contribution to 
the gauge coupling ${\rm Re} S$ and 1-loop ${\rm Re} T$ is necessary 
for the approximate GUT relation. We have seen, however, that 
the 1-loop ${\rm Re} T$ to the visible sector and $-{\rm Re} T$ 
to the hidden sector corresponds to the warped factor in the 
11-th direction in language of Heterotic M-theory. The warped 
metric is a consequence of asymmetric distribution of the second 
Chern class (instanton numbers). Once we have such a geometric 
meaning, 
\begin{equation}
 \frac{\alpha_{\rm hidden}}{\alpha_{\rm GUT}} = 
 \frac{{\rm vol}(Z)|_{x_{11}=\pi \rho}}{{\rm vol}(Z)|_{x_{11}=0}},
\end{equation}
and a hierarchy is easily generated between the two gauge coupling 
constants, unless the ``instanton numbers'' are distributed precisely 
the same in the visible and hidden sectors. Thus, actually 
the approximate GUT relation does not require a fine-tuning;
we can understand it as a natural consequence of dynamics of 
Ramond--Ramond field and metric in the 11th direction.

This is still a predictive framework of GUT. Conventional 
unified theories use two continuous parameters, $M_{\rm GUT}$ and 
$\alpha_{\rm GUT}$, to fit two gauge coupling constants, e.g., 
$\alpha_C$ and $\alpha_Y$, and predict the last one, e.g., $\alpha_L$. 
Now, in this framework, three continuous parameters are involved, 
namely, $\kappa^2$, $\rho$ and the compactification scale 
${\rm vol}(Z)|_{x_{11}=0}$, but there are four observable data
that are given by those parameters, namely the three gauge coupling 
constants $\alpha_{C,L}$, and $\alpha_{\tilde{Y}}$, and the Planck scale. 
When the three parameters are use to fit $\alpha_{C,L}$ and the Planck 
scale, this framework predicts that $\alpha_{\tilde{Y}}$ is quite close 
to $\alpha_{C,L}$ at the unification scale, and is a little smaller. 
We know that this prediction is consistent with the precise measurement 
of the Standard Model gauge couplings at LEP. 
See Figure~\ref{fig:unif}.

Note that it is not necessary to assume (\ref{eq:Theta}) for the 
disparity between the gauge coupling constants of unbroken non-Abelian 
symmetries in the visible and hidden sectors; the running of 
${\rm vol}(Z)$ along the $x_{11}$ direction is always given by 
$\propto (1 - \alpha/8\sqrt{2} x_{11})^2$, whether (\ref{eq:Theta}) 
is satisfied or not. 
However, we keep this assumption because we need another phenomenological 
requirement, namely $A \propto \int_Z c_1(L_Y)^2 \wedge J = 0$. 
As one can see from (\ref{eq:def-T}--\ref{eq:def-A}), $A$ can 
potentially be of order of $T$. Even if warped metric 
in the $x_{11}$ direction accounts for why 
${\rm Re}(S-T) \ll {\rm Re}(S+T)$, non-vanishing 
$A \approx {\cal O}(S,T)$ in the kinetic mixing matrix 
(\ref{eq:gauge-kine-matrix}) invalidates the scenario in this section.
The K\"{a}hler form is expanded as in (\ref{eq:JB}), and the coefficients 
$\alpha^k(x_{11})$ would run differently in the $x_{11}$ direction, if 
(\ref{eq:Theta}) were not satisfied. If $\alpha^k$'s change their 
ratio among them over the interval $x_{11} \in \left[ 0, \pi \rho \right]$, 
then $A$ will not vanish even if it does somewhere in the interval.
Thus, in order to impose that $A=0$, we assume (\ref{eq:Theta}). 

This may not be a problem because $A$ is of order $\kappa^{\frac{2}{3}}$ 
to begin with, and the running effect of $A$ in $x_{11}$ comes only in 
another $\kappa^{\frac{2}{3}}$ order, hence in the next-to-next-to-leading 
order, ${\cal O}(\kappa^{\frac{4}{3}})$. But, for making an error in
safe side,\footnote{$A=0$ when the volume of certain cycle vanishes, 
as we discuss later. In this sufficient condition for $A=0$, 
some K\"{a}hler moduli are chosen to be zero. If the running of 
$\alpha^k$ is totally arbitrary, as oppose to the case (\ref{eq:Theta})
when $\partial_{11} \alpha^k \propto \alpha^k$, some of $\alpha^k$, 
already chosen to be zero may run into negative value. The Heterotic 
M theory compactification in this case is geometric in part of 
the interval of $x_{11} \in \left[ 0,\pi \rho \right]$, while 
possibly non-geometric for the rest of the interval. Such a situation 
is avoided when (\ref{eq:Theta}) is satisfied.} 
as well as for simplicity, we maintain the assumption 
(\ref{eq:Theta}) in what follows. 

\subsubsection{Generalized Green--Schwarz Coupling in the Heterotic-M Theory}

Just like $\alpha^k$, the coefficients of the K\"{a}hler form, 
run in $x_{11}$ when (\ref{eq:asymmetry}) does not vanish, 
the zero modes from the Ramond--Ramond 3-form field $C^{(3)}$, i.e., 
$b^k$ in (\ref{eq:JB}), also have non-trivial wavefunction along 
the $x_{11}$ direction \cite{SW}. Thus, one has to check whether 
the generalized Green--Schwarz coupling (\ref{eq:gen-GS}) of $D=4$ 
effective theory is modified or not; the discussion so far on 
the gauge coupling unification is based on an assumption that 
only the gauge coupling constants $1/g^2$ 
and $1/g^{'2}$ are affected by the warping geometry, but the 
linear combination coefficients of the generalized Green--Schwarz 
coupling (\ref{eq:gen-GS}) are not.

It is sufficient to see the coefficients of the the cross terms of 
(\ref{eq:gen-GS}), now in the warped compactification of the 
Heterotic M theory. The cross term originates from the interaction 
\begin{equation}
 \frac{1}{2\sqrt{2} \pi \kappa^2} \left(\frac{\kappa}{4\pi} \right)^{\frac{2}{3}}
  \int_{11D} \widetilde{C}^{(6)} \wedge \left( J_1 \delta(x_{11}) 
+  J_2 \delta(x_{11}-\pi \rho) \right),
\label{eq:bdry}
\end{equation}
where 
\begin{equation}
J_1 = \tr {}_1 \left(\frac{F}{2\pi}\right)^2 - \frac{1}{2} 
 \tr \left(\frac{R}{2\pi}\right)^2, \qquad 
J_2 = \tr {}_2 \left(\frac{F}{2\pi}\right)^2 - \frac{1}{2} 
 \tr \left(\frac{R}{2\pi}\right)^2.
\end{equation}
$\widetilde{C}^{(6)}$ is related to $C^{(3)}$ via 
$d \widetilde{C}^{(6)} = *_{11D} d C^{(3)}$.
The interaction above yields the source term to the Bianchi identities 
\begin{equation}
d G^{(4)} = - \frac{1}{2\sqrt{2} \pi} 
    \left(\frac{\kappa}{4\pi}\right)^{\frac{2}{3}}
    \left(\delta(x_{11}) J_1 + \delta(x_{11}-\pi \rho) J_2 \right).
\end{equation}

The wavefunction of the zero modes from $C^{(3)}$ have the form \cite{SW} 
\begin{equation}
 C^{(3)} = \omega_k \wedge d x_{11} e^{f(x_{11})/2} b^k(x^\mu) + \cdots.
\label{eq:C3-0mode}
\end{equation}
Here, we maintained only the modes in the chiral multiplets $T^k$, 
dropping the one in $S$, because $Q^S=0$ and we are interested in 
the generalized Green--Schwarz interaction involving the K\"{a}hler 
moduli chiral multiplets.
Now, we take the Hodge dual of this zero-mode wavefunctions.
They are 
\begin{equation}
d \widetilde{C}^{(6)} = 
 \left( \epsilon_{\mu \nu \lambda \kappa} \partial^\mu b^k(x) dx^\nu dx^\lambda dx^\kappa
 \right) \wedge 
 ( *_6 \omega_k ) + \cdots ,
\end{equation}
where $*_6$ is the Hodge dual on a Calabi--Yau 3-fold $Z$ with the 
unwarped K\"{a}hler metric $g_{\alpha \bar{\beta}}$. 
The warped factor $e^{f(x_{11})/2}$ in (\ref{eq:C3-0mode}) is cancelled 
and disappears in $\widetilde{C}^{(6)}$ after taking the Hodge dual. 
Thus, the coefficients of the cross term in (\ref{eq:gen-GS}), 
which arises from (\ref{eq:bdry}), are not suppressed or enhanced 
by the warped factor $e^{f(x_{11})}$. Therefore, the discussion until 
section 3.2.1 does not have to be altered.

\subsection{Phenomenological Aspects}

\subsubsection{Fayet--Iliopoulos Parameters and a Global U(1) Symmetry}

Let us take a brief look at Fayet--Iliopoulos parameters of 
those U(1) symmetries. They are given by 
\begin{eqnarray}
\xi_Y & = & \tr ({\bf q}_Y^2) \frac{M_G^2}{32\pi^2} 
  \left(\frac{2\pi l_s^2}{{\rm vol}(Z)} \int_Z c_1(L_Y) \wedge J \wedge J
        - \frac{g_{\rm YM}^2}{2} e^{2\tilde{\phi_4}}Q^S_Y
  \right), \\
\xi_2 & = & \tr ({\bf q}_2^2) \frac{M_G^2}{32\pi^2} 
  \left(\frac{2\pi l_s^2}{{\rm vol}(Z)} \int_Z c_1(L_Y) \wedge J \wedge J
        - \frac{g_{\rm YM}^2}{2} e^{2\tilde{\phi_4}}Q^S_2
  \right),
\end{eqnarray}
where $e^{- 2\tilde{\phi}_4} = e^{- 2 \tilde{\phi}} 
{\rm vol}(Z)/\vev{{\rm vol}(Z)}$, 
and they enter in the $D=4$ effective theory as 
\begin{equation}
{\cal L} = - \frac{\tr ({\bf q}_Y^2)}{2g^2}D_Y^2 
 - \frac{\tr ({\bf q}_2^2)}{2g^{'2}} D_2^2 
 + D_Y \left(\xi_Y + {\bf q}_Y \phi^\dagger \phi \right)
 + D_2 \xi_2.
\end{equation}
The auxiliary fields $D_Y$ and $D_2$ are rotated just as 
the vector fields $A_Y$ and $A_2$ are, and the Fayet--Iliopoulos 
parameters are also re-organized accordingly. Thus, 
Fayet--Iliopoulos parameters of the U(1)$_{\rm massive}$ and 
$\U(1)_{\tilde{Y}}$ vector multiplets are given by linear 
combination of $\xi_Y$ and $\xi_2$. 

Zero modes from the visible sector---denoted by $\phi$ above---carry 
charges under the massless $\U(1)_{\tilde{Y}}$ and massive $\U(1)$, and 
if there are zero modes from the hidden sector charged under the 
$\U(1)_2$ symmetry, then they are also charged under the both. 
If the Fayet--Iliopoulos parameter of the massive $\U(1)$ does 
not vanish, and if it is absorbed by vev's of chiral multiplets, 
then their vev's break the $\U(1)_{\tilde{Y}}$ symmetry as well. 
Thus, the Fayet--Iliopoulos parameters of both $\U(1)_{\rm massive}$ 
and $\U(1)_{\tilde{Y}}$ have to vanish, and so do $\xi_Y$ and $\xi_2$ 
(at the supersymmetric limit).

Geometry of Calabi--Yau 3-fold and vector bundles on it has to be 
arranged so that just the matter spectrum of the supersymmetric 
standard model arise from the visible sector. Thus, the 
$\U(1)_Y\left[\SU(3)_C\right]^2$ and 
$\U(1)_Y\left[ \SU(2)_L\right]^2$ mixed anomalies vanish.
It is known that the coefficient of the one-loop Fayet--Iliopoulos
parameters $Q^S$ of (possibly anomalous) U(1) symmetries are 
proportional to the U(1)-[non-Abelian]$^2$ mixed anomaly in 
low-energy effective theories of the Heterotic $E_8 \times E_8'$ 
string theory,\footnote{\label{fn:E8-SO32}
Reference \cite{Polchinski} argues based on field theory that 
1-loop Fayet--Iliopoulos parameters are proportional to 
U(1)-[gravity]$^2$ anomalies of low-energy spectrum, but this 
argument implicitly assumes that quadratically divergent contributions 
from any one of massless chiral multiplets are regularized exactly 
in the same way. It is very subtle, however, to discuss cancellation 
among divergent quantities, and it is more appropriate to study 
this issue (Fayet--Iliopoulos parameter) in a UV finite framework 
such as string theory. 
In a compactification of Heterotic $\SO(32)$ string theory with 
an SU(3) vector bundle, Fayet--Iliopoulos parameter of a U(1) 
vector multiplet was calculated explicitly, and it turned out 
to be proportional to U(1)-[gravity]$^2$ indeed \cite{ADS, DIS}.
Reference~\cite{MunichSO(32)} further showed that this is true 
for Calabi--Yau 3-fold compactifications of Heterotic $\underline{\SO(32)}$ 
string theory with generic (supersymmetry preserving) vector 
bundles. In compactifications of Heterotic $\underline{E_8 \times E'_8}$
string theory, however, \cite{Munich0504} showed that the 1-loop 
Fayet--Iliopulos parameters $Q^S$ are proportional to 
U(1)-[non-Abelian]$^2$ mixed anomalies. $Q^S$ does not have to 
be proprotional to U(1)-[gravity]$^2$ anomalies, because various 
massless multiplets originate from cohomology groups of vector 
bundles in various representations, and UV divergent contriubtions 
to Fayet--Iliopoulos parameters from those multiplets are not
regularized (cut-off and made UV-finite) exactly in the same way. \\
Section 3 of \cite{ADS} argues, however, that the 1-loop
Fayet--Iliopoulos parameters (i.e. $Q^S$) are proportional to 
U(1)-[gravity]$^2$ anomalies in compactifications of 
Heterotic $E_8 \times E'_8$ string theory as well. 
We have not yet clarified how the two apparently contradicting 
statements from \cite{ADS} and \cite{Munich0504} are related.
In this article, we adopted the statement in \cite{Munich0504}.
} 
and hence $Q^S$ vanishes for $\xi_Y$. 
Without the 1-loop term, the tree-level term should also vanish 
in order for $\xi_Y$ to vanish. Thus, 
\begin{equation}
\int_Z c_1(L_Y) \wedge J \wedge J = 0.
\end{equation}
It also follows from this condition that $Q_2^S=0$ by requiring $\xi_2=0$. 
All of this argument ignores all the non-perturbative (and stringy) 
corrections to the Fayet--Iliopoulos parameters.

\subsubsection{Orbifold GUT and Beyond}

{\bf Localized $\U(1)_Y$ Breaking} \\

Two assumptions that are essential in maintaining the gauge coupling 
unification are 
\begin{equation}
 \int_Z J \wedge J \wedge c_1(L) = 0, \qquad \int_Z J \wedge c_1(L)^2 = 0.
\label{eq:U(1)-break-constraint}
\end{equation}
The first one comes from the stability condition of the vector bundle 
$V_1$ (also from requiring the vanishing Fayet--Iliopoulos parameters 
$\xi_{2,Y}$), and the second one was introduced right after (\ref{eq:def-A})
in order to bring the 1-loop threshold corrections under control.
These conditions are derived in the supersymmetric and large-volume 
limit.

Suppose that $c_1(L_Y)$ is given by 
\begin{equation}
c_1(L_Y) = \sum_I n_I D_I,
\label{eq:c1-LY-def}
\end{equation}
where $D_I$ are divisors of a Calabi--Yau 3-fold $Z$, and 
$n_I$ coefficients. 
The first equation of (\ref{eq:U(1)-break-constraint}) becomes
\begin{equation}
 \int_Z J^2 \wedge (\sum_I n_I D_I) = \sum_I n_I \int_{D_I} J^2 = 0.
\end{equation}
This condition is satisfied, if all the $D_I$'s that appear in 
(\ref{eq:c1-LY-def}) have vanishing sizes, for example. 

The second equation of (\ref{eq:U(1)-break-constraint}) becomes 
\begin{equation}
\int_Z J \wedge c_1(L)^2 = \sum_I \sum_J n_I n_J \int_{D_I \cdot D_J} J = 0.
\end{equation}
If all the curves $D_I \cdot D_J \neq \phi$ have vanishing volumes, 
then the second condition is also satisfied. 

For an example, $T^6/\Z_3$ orbifold has 27 isolated vanishing 
exceptional divisors, each of which is isomorphic to $\C P^2$.
Another example is $WP_{1,1,1,3,3} \supset (9)$, which also contains 
3 isolated $\C^3/\Z_3$ singularities, and hence 3 such divisors 
each of which is isomorphic to $\C P^2$.
Reference \cite{HN} argued that containing a source of SU(5)$_{\rm GUT}$ 
symmetry breaking into an orbifold singularity brings the threshold 
correction under control. Indeed, we found that the 
1-loop threshold corrections to the U(1)$_Y$ gauge coupling is proportional 
to $A$, and hence this correction is made small when $A=0$ [stringy 
correction would remain, but it will not have a large-volume enhancement]. 
Thus, we largely confirm their claim that the 1-loop threshold correction 
can be made small when the symmetry breaking is confined to orbifold 
singularities. 
By now, we see that (\ref{eq:U(1)-break-constraint}) is the generalized 
version of the idea of \cite{HN}, and it is obvious that the global 
geometry does not have to be a toroidal orbifold, as long as 
(\ref{eq:U(1)-break-constraint}) are satisfied. This generalization 
should allow much more variety in the choice of geometry. 

{\bf Naive Dimensional Analysis}

There are a couple of different sources that give rise to a 
small deviation from the GUT relation. As we have seen, one 
of such sources was the mixing with an extra massless 
strongly coupled gauge field. The extra contribution to the 
gauge coupling $(3/5)/g_{\tilde{Y}}^2$ is suppressed relatively to the 
leading contribution $\simeq 1/g^2_C \simeq 1/g_L^2$ by a factor 
of order 
\begin{equation}
  \frac{{\rm vol}(Z)|_{x_{11} = \pi \rho}}{{\rm vol}(Z)|_{x_{11} = 0}} 
   \simgt \frac{\alpha^{'3}}{{\rm vol}(Z)|_{x_{11} = 0}}.
\label{eq:tree-order}
\end{equation}
Since the observed values of the Planck scale, GUT scale and 
the unified gauge coupling constant suggest that the 
${\rm vol}(Z)|_{x_{11} = \pi \rho}$ is almost close to 
$\alpha^{'3}$ \cite{WittenStrongGs, CK1, CK2}, the inequality 
above is almost saturated in the reality, and it can be quite small. 

Only supergravity approximation (large-volume limit) was used 
in (\ref{eq:gauge-kine-matrix}) in the expression 
for the threshold corrections to the gauge coupling constants.
There will be extra stringy contributions, which cannot be 
captured by supergravity approximation. Since there are literatures 
on the threshold corrections to the gauge kinetic functions, 
results in such references can be used to obtain a precise 
estimate of how large they are (to the level of whether some power 
of $\pi$ is involved or not). This article does not cover such 
calculation, however. Instead, we just assume in this article 
that they are of order unity, because there is no characteristic 
scales other than the string scale for such contributions.  
Since we consider a situation where $\Re S \sim \Re T \sim R^2/\alpha'$, 
the order-one stringy and possibly $\SU(5)_{\rm GUT}$-breaking 
corrections to the gauge coupling are relatively 
\begin{equation}
 \frac{{\cal O}(1)}{\Re T} \sim  \frac{\alpha'}{R^2},
\end{equation}  
compared with the leading term $\Re (S+T)$. 
Therefore, this correction is more important than 
(\ref{eq:tree-order}), which may be of order 
${\cal O}((\alpha'/R^2)^3)$. Orbifold calculations 
may be useful, as we mentioned above, in obtaining more precise 
estimate of the stringy corrections to the GUT relation.

\subsubsection{Dimension-6 Proton Decay}

Here is a remark on dimension-6 proton decay. 
As for the process of determining the Kaluza--Klein scale 
from observables, we do not have much to add to what we 
already wrote at the end of section 2.2. The dimension-6 
proton decay operators are generated after massive gauge 
bosons. Two vertices, each of which involves two fermion 
zero modes and one massive gauge boson, are combined together.
The three-point vertex comes from the covariant derivative interaction 
of the gaugino kinetic term. Quarks and leptons come from 
a part of gaugino in the adjoint representation of $E_8$, 
and Kaluza--Klein tower of off-diagonal gauge bosons 
in $\SU(5)_{\rm GUT}$ is also a part of $E_8$ gauge filed 
on 10 dimensions. The coefficient of the three-point vertex 
is calculated by overlap integration over the Calabi--Yau 3-fold 
for the compactification. 

In toroidal orbifold compactification, (fermion) zero-mode 
from untwisted sector (bulk) has absolutely flat wavefunctions,
while that of the Kaluza--Klein gauge bosons are Fourier modes 
on the torus. The overlap integration involving two untwisted-sector 
fermions and one Kaluza--Klein gauge boson vanishes, and 
the Kaluza--Klein gauge bosons do not induce a transition between 
zero-mode fermions from the untwisted sectors. Although such 
predictions appear in the literature from phenomenology community, 
they should hold only for toroidal orbifold compactifications. In general 
Calabi--Yau 3-fold compactification of the Heterotic string theory, 
wavefunctions of zero-modes of chiral multiplets are identified with 
elements of bundle-valued cohomology groups on a Calabi--Yau 3-fold, 
and they are not absolutely flat on a curved manifold.
Products of two cohomology group elements multiplied by a higher 
harmonic function do not vanish generically, after being integrated 
over a Calabi--Yau manifold. Branching fractions of various 
decay modes of a proton can be generation dependent, but 
more detailed geometric data is necessary in order to calculate 
branching fractions for individual models of Heterotic string 
compactification.

\subsection{Digression: Landscape of Unified Theories}

Our presentation has consisted in considering the Georgi--Glashow
SU(5)$_{\rm GUT}$ unified theories and study how to break
the $\SU(5)_{\rm GUT}$ symmetry down to the Standard-Model
$\SU(3)_C \times \SU(2)_L \times \U(1)_Y$. There are
other types of unified theories, among which flipped SU(5) model 
and Patti--Salam model will be the most famous. 
We could have studied how to construct such unified theories, and
then consider how to break those unified symmetries.

Our choice of Georgi--Glashow $\SU(5)_{\rm GUT}$ is not without a reason.
The electroweak mixing angles in the quark sector are all small,
but those in the lepton sector are large (apart from the last one
yet to be measured). In Pati--Salam type unified theories, the
quark doublets and lepton doublets are contained in a common
irreducible representation of the unified gauge group. In order to
obtain the qualitative pattern of the electroweak mixing stated
above, one generically needs to have Yukawa couplings that
heavily involve the source of symmetry breaking of the Pati--Salam
gauge group. In the flipped SU(5) model with Froggatt--Nielsen
(or Abelian flavour symmetry) type Yukawa matrices,\footnote{It is known 
that Yukawa couplings follow such a pattern in certain region of the 
moduli space; examples include small torus fibered compactification 
\cite{HSW} and near-orbifold-limit region.} not all the
Yukawa eigenvalues and mixing angles come out right either, meaning
presumably that the Yukawa couplings heavily involve symmetry breaking
of the flipped SU(5) symmetry. The Georgi--Glashow $\SU(5)_{\rm GUT}$
symmetry does not have this problem, and it can be a fairly well
approximate symmetry  (to some extent)\footnote{In most generic 
compactification of Heterotic string theory, it may not be an important 
issue whether or not $\SU(5)_{\rm GUT}$ (or some other unified-theory 
gauge group) is a good approximate symmetry with respect to Yukawa couplings. 
Gauge field background is introduced in the $\U(1)_Y$ direction, and 
zero modes (cohomology group elements) in the same irreducible 
representation of $\SU(5)_{\rm GUT}$ but with different $\U(1)_Y$ charges 
have different zero-mode wavefunctions, and hence the Yukawa couplings 
are not expected to satisfy relations that would have followed from 
$\SU(5)_{\rm GUT}$ symmetry. In Heterotic vacua with F-theory dual  
and also in F-theory vacua, however, zero modes from a common $\SU(5)_{\rm GUT}$ 
representation are localized in a common matter curve, and flavour 
properties associated with fields in representations such as {\bf 10} or 
$\bar{\bf 5}$ may be attributed to some properties of matter curves of 
corresponding representations. Thus, it is not a meaningless question 
which unified symmetry is a good approximation in Yukawa matrices. 
A relevant discussion is found in \cite{HSW}.} 
in Yukawa couplings of quarks and leptons.

In field-theory model building, different types of unified theories 
are just different models. It is a matter of which model provides 
better approximation to the reality. 
From the perspective of (landscape of) string theory, however, 
things begin to look a little different.  
If the moduli space of various Calabi--Yau manifolds and vector
bundles are interconnected,\footnote{Note, however, 
that we are not interested in dynamical (or cosmological) transitions 
between vacua in this article. Thus, we are not concerned about 
whether there is a topological barriers within the moduli space. 
Note also that the connectedness of landscape of vacua depends on the 
``sea level''---how much symmetry breaking one allows when one goes 
from one vacuum to another.
} 
there may not actually be a definite distinction between various types of 
unified theories. 
From one vacuum in one type of unified theory to another in a different
type of unified theory, it may be possible to deform continuously
over the moduli space (before introducing fluxes). 
Low-energy observables such as Yukawa eigenvalues and mixing angles 
are functions of moduli, and they change continuously until they 
look phenomenologically qualitatively different. 
Thus, any types of unified theories in landscape of string vacua 
cannot be absolutely ``wrong''; it is just a matter of how far those vacua 
are from ours. String landscape accommodates hundreds of models of 
unified theories, and may set a stage to discuss dynamical selection 
of models of unified theories. String landscape works as 
a unified theory of unified theories.

In what follows, we study the relation between Georgi--Glashow SU(5) 
and flipped SU(5) unified theories in string landscape. We will be 
very crude in that we do not restrict ourselves to a partial moduli 
space where matter parity is preserved, or to a moduli space where 
vector bundles have appropriate extension structure.

Both the Georgi--Glashow SU(5) gauge group and the flipped
SU(5)$'$ gauge group can be embedded in a common SO(10) model.
Thus, it is easiest to see how those theories are obtained 
by breaking $\SO(10)$ symmetry. Georgi--Glashow SU(5)$_{\rm GUT}$ symmetry 
is the commutant of a $\U(1)_\chi$ in a maximal torus, specified by 
a charge vector ${\bf q}_\chi$ in the Cartan subalgebra. The gauge group 
of the flipped SU(5), $\SU(5)' \times \U(1)_{\chi'}$ is the commutant 
of $\U(1)_{\chi'}$ generated by ${\bf q}'_\chi$. Those two theories 
share a rank-5 Cartan subalgebra of $\SO(10)$, and the charge vectors 
are related by  
\begin{eqnarray}
{\bf q}_\chi = \diag (2,2,2,2,2), & \qquad & {\bf q}_Y = \diag \left(
   -\frac{1}{3},-\frac{1}{3},-\frac{1}{3},\frac{1}{2},\frac{1}{2} \right), \\
{\bf q}'_\chi = \diag (2,2,2,-2,-2), & \qquad & {\bf q}'_Y = \diag \left(
   -\frac{1}{3},-\frac{1}{3},-\frac{1}{3},-\frac{1}{2},-\frac{1}{2} \right).
\end{eqnarray}
Those charge vectors satisfy 
\begin{equation}
 \left( \begin{array}{c}
   {\bf q}'_{\chi} \\ {\bf q}'_{Y}
 \end{array} \right) = \frac{1}{5} 
 \left( \begin{array}{cc} 
   1 & - 24 \\ -1 & -1 
 \end{array} \right)
 \left( \begin{array}{c}
   {\bf q}_{\chi} \\ {\bf q}_{Y}
 \end{array} \right). 
\end{equation}

In order to obtain Georgi--Glashow SU(5) unified theories 
in compactification of the Heterotic $E_8 \times E_8$ string theory, 
we can begin with an $\SU(4)$ vector bundle $V_4$ and a line bundle 
$L_{\chi}$. By further turning on vev's in zero modes 
$H^1(Z; V_4 \otimes L_{\chi}^{-5})$ and
$H^1(Z; V_4^\times \otimes L_{\chi}^5)$, one obtains an $\SU(5)$ bundle,
leaving unbroken Georgi--Glashow $\SU(5)_{\rm GUT}$ symmetry.
Vev's in the zero modes are regarded as deformation of the vector bundle, 
since those cohomology groups describe the deformation of the bundles.
The Georgi--Glashow SU(5)$_{\rm GUT}$ symmetry can be broken down to 
$\SU(3)_C \times \SU(2)_L$ (and $\U(1)_Y$) when a line bundle $L_Y$ 
is turned on in the direction specified by ${\bf q}_Y$.

The flipped SU(5) theories are obtained in Heterotic string 
compactification\footnote{
 In the flipped SU(5) unified theories, one needs to assume that
 the gauge coupling constant of $U(1)_{\chi '}$ is the same as that of
 $\SU(5)'$ in order to obtain the GUT relation after the symmetry breaking
 due to the vev. This assumption seems to be satisfied when they are
 obtained through compactification of string theory containing
 SO(10) gauge group, because the $\U(1)_{\chi '}$ symmetry originates
 from the same SO(10) gauge group. However, a line bundle in the
 $\U(1)_{\chi '}$ direction removes the massless $\U(1)_{\chi'}$ gauge
 field from the spectrum, just like in the case of $\U(1)_Y$ gauge field.
 Thus, an extra gauge field has to be obtained through a line bundle
 sharing the same first Chern class with $L_{\chi'}$. In order to
 maintain the approximate GUT relation, the gauge coupling of the 
combined massless U(1) gauge field should be almost the same 
as that of $\U(1)_{\chi '}$. This is achieved when the extra U(1) 
gauge field has a large coupling constant, just like in sections 2 and 3.
 The same idea works for the flipped SU(5) unified theories as well.}
by turning on the same $\SU(4)$ bundle $V_4$ and a line bundle
$L_{\chi'}$ in the direction specified by ${\bf q}'_{\chi}$. 
Furthermore, vev's are turned on within
zero modes ${\bf 10}'$'s $=H^1(Z; V_4 \otimes L_{\chi'}^{-1} \otimes L_{Y'}^{+1})$ 
and its conjugate, so that the $\SU(5)' \times \U(1)_{\chi'}$ symmetry 
is broken down to $\SU(3)_C \times \SU(2)_L \times \U(1)_Y$. 
($L_{Y'}$ is a trivial bundle in the flipped SU(5) models; we included 
$L_{Y'}$ in the expression above to clarify where the chiral multiplets 
in the ${\bf 10}'$ representation vev's should develop.)
Vev's in these chiral multiplets correspond to deformation
of the vector bundle. The structure group is enlarged. 

Because of the relation among the charge vectors above, we find 
a translation 
\begin{equation}
 L_\chi \leftrightarrow L_{\chi'}^{\frac{1}{5}} \otimes L_{Y'}^{-\frac{1}{5}}, \qquad 
 L_Y \leftrightarrow L_{\chi'}^{-\frac{24}{5}} \otimes L_{Y'}^{-\frac{1}{5}}.
\end{equation}
Thus, the deformation of the bundle in the flipped SU(5) unified theories 
$H^1(Z; V_4 \otimes L_{\chi'}^{-1} \otimes L_{Y'})$ is actually the same 
deformation as the one in Georgi--Glashow SU(5) unified theories, 
$H^1(Z; V_4 \otimes L_\chi^{-5})$.
When one talks of flipped SU(5) unified theories, one usually
assumes that the Kaluza--Klein scale is higher than the unification
scale, where the vev in ${\bf 10}'$ breaks the symmetry. 
As the vev increases, and it becomes comparable to the Kaluza--Klein 
scale, however, it is more appropriate to treat the vev as 
a part of vector bundle moduli.
In the large vev limit of the flipped SU(5) unified theories, 
a rank-5 vector bundle breaks $\SU(5) \subset E_8$ containing 
$\SU(4)$ and $\U(1)_\chi$, and a line bundle still remains, 
with the structure group of the U(1) bundle set in the direction 
\begin{equation}
 {\bf q}'_\chi  \equiv \frac{1}{5} {\bf q}_Y \qquad ({\rm mod~} {\bf q}_\chi).
\end{equation}
Thus, this is nothing but the Georgi--Glashow SU(5) unified theories 
with a line bundle in the $\U(1)_Y$ direction. 


\section{F-theory Vacua}
\label{sec:F}

The $\SU(5)_{\rm GUT}$ symmetry can be broken by turning on
a line bundle in the $\U(1)_Y$ direction. The line bundle
is given by a 2-form field strength tensor of a gauge
field on the D7-brane world volume in the perturbative
Type IIB string theory, and in F-theory vacua in general,
essentially the same thing is expressed by a four-form field
strength borrowing language of M-theory.

The $\U(1)_Y$ problem exists for such models, just like
we already explained in section~\ref{ssec:Y-problem}
in Type IIB models, and the Green--Schwarz coupling that
makes the $\U(1)_Y$ gauge field a mass term is rephrased
from the Chern--Simons interaction on the D7-brane worldvolume
in the Type IIB description to the Chern--Simons term
in the eleven-dimensional supergravity.

The $\U(1)_Y$ problem can be, in principle, solved by allowing
an extra U(1) gauge symmetry to mix with the U(1)$_Y$ gauge
field contained in the SU(5)$_{\rm GUT}$ symmetry; the extra U(1)
gauge field has to be strongly coupled so that the deviation
from the GUT relation is not too large.
Note that the unification between the $\SU(2)_L$ and $\SU(3)_C$
gauge coupling constants is already achieved, by wrapping two and
three D7-branes (or by just having a locus of $A_4$ singularity)
on a common holomorphic 4-cycle.

The extra U(1) gauge field may arise from an extra D7-brane
(or from an extra 7-brane locus in F-theory in general).
In order to obtain a little hierarchy between the GUT scale
(Kaluza--Klein scale) and the Planck scale of the $D=4$ effective
theory, the volume of $A_4$ singularity is chosen to be
parametrically large in string scale units. Since the gauge kinetic
function $1/g^2$ is roughly proportional to the volume of the 4-cycle
a D7-brane is wrapped in the Type IIB string theory, the unified
gauge coupling constant $1/g^2_C \sim 1/g^2_L$ is small.
An effective theory below the Kaluza--Klein scale becomes
perturbative, just like we expect the MSSM to be.
On the other hand, if the extra 7-brane is wrapped on a 4-cycle
whose volume is of order one in string scale units, its
gauge kinetic function remains small, and the gauge theory
on the 7-brane is strongly coupled. Thus, the deviation of the
$\U(1)_{\tilde{Y}}$ coupling from the GUT relation is (positive, in
$\Delta (1/g^2)$, and) small as long as the extra U(1) gauge theory
is strongly coupled. This picture dates back to \cite{WY} (and further
to \cite{IWY}), where fractional D3-branes were used as
the extra 7-brane; fractional D3-braes are known to be wrapped
(possibly anti-) D7-branes or D5-branes depending on a nature
of singularity. Why some 4-cycle has a parametrically large volume,
and some do not is a question associated with stabilization of
K\"{a}hler moduli. Thus, the $\U(1)_Y$ problem is translated into
a problem of moduli stabilization.
Since the doublet and triplet part of the Higgs multiplets are
regarded as global holomorphic sections of different line bundles
(they differ by $L_Y^{\otimes 5}$), the massless spectrum of doublets
and triplets can be different, giving a solution to the
doublet--triplet splitting problem.

There may be threshold corrections to the gauge kinetic functions
of order
\begin{equation}
  \int_\Sigma J \wedge c_1(L_Y), \qquad \int_\Sigma c_1(L_Y)^2.
\end{equation}
The first term should vanish because it is the stability condition
(or the Fayet--Iliopoulos D-term parameter of the
$\U(1)_Y$ symmetry). There may be a threshold corrections of the
order of the second term above to the gauge coupling $1/g^2_Y$,
but it is small by a factor of $\alpha^{'2}/{\rm vol}(\Sigma) \sim
\alpha^{'2}/R^4$ compared with the leading order term.
Thus, the threshold correction does not affect the GUT
relation seriously.

As we discussed at the end of section 2.2, dimension-6 proton decay 
is expected to be fast. Furthermore, in F-theory vacua, there 
may be an extra enhancement in the decay rate, because 
the amplitude receives an UV-divergent enhancement factor 
when matter multiplets are localized in the extra dimensions 
\cite{friedmanwitten}.
The enhancement factor depends on the number of codimensions in which 
matter multiplets are localized relatively to gauge fields, and 
if there is an UV-divergent factor indeed, then string theory calculation 
has to be involved in making an estimate of the form factor, just like 
in \cite{klebanovwitten}. It is an interesting open problem what the 
enhancement factor will be in F-theory models.

{\bf Note added in version 3:} 
This enhancement factor was studied in \cite{DW2} after the first
version of this article. The prediction of the enhancement factor 
is indeed one of the most important results of F-theory 
phenomenology; it is not mere a hindsight explanation of 
known parameters of the Standard Model, but it does change 
the prediction of observables in experiments in the future (i.e., 
that is physics!). 
Furthermore, the enhancement factor directly originates from 
localization of quarks and leptons in internal dimensions relatively 
to gauge fields, and hence is clearly an effect that is absent 
in field theory models purely in 3+1 dimensions (i.e., that is 
string theory!). 
Note also that the enhancement factor is very robust, 
in that it does not depend on details of how $\SU(5)_{\rm GUT}$ 
symmetry is broken. It is (and the following result is) applied 
to the $\SU(5)_{\rm GUT}$ breaking scenario discussed in this article, 
as well as to the scenario in case of non-surjective pull-back of 
2-forms, $i^*: H^2(B_3) \rightarrow H^2(S)$, in \cite{Buican, Vafa2, DW2}. 
Unfortunately the paper \cite{DW2} considered that dominant
contribution to (\ref{eq:86}) and (\ref{eq:91}) comes from a subset where
the Green function $G_{int}$ diverges, but this subset
is measure zero. In this version of our article we present
our study on the enhacement factor by evaluating (\ref{eq:86}) and
(\ref{eq:91}) carefully, and we obtain predictions on dimension-6
proton decay that are qualitatively different from those of \cite{DW2}.

Before we begin to study the enhancement factor in F-theory
compactifications, let us briefly review the essence 
of \cite{friedmanwitten} in M-theory compactifications on manifolds 
with $G_2$ holonomy. In $G_2$ holonomy compactifications with $\SU(5)$ 
unification, $\SU(5)_{\rm GUT}$ gauge fields propage on a 3-cycle $Q$, 
and charged matter fields such as those in ${\bf 10}$ and 
$\bar{\bf 5}$ representations are localized at isolated points 
$\vec{y}_i$ in $Q$; here, we use $\vec{y}$ as coordinates of $Q$, and 
$x^\mu$ for the Minkowski space $\R^{3,1}$. Charged matter fields 
are coupled to the gauge field as 
\begin{equation}
 \int_{\R^{3,1}} \!\! d^4 x \; \; J^\mu_i(x) A_\mu(x,\vec{y}_i),
\end{equation}
and the gauge field has a kinetic term 
\begin{equation}
 - \frac{1}{4 g_7^2}\int_{\R^{3,1}} d^4 x \int_Q d^3y \; \; 
   \tr (F_{MN} F^{MN}).
\end{equation}
The gauge field $A_M$ and current $J^\mu$ are given 1 and 3 mass
dimensions, respectively, and the gauge coupling $g_7^2$ have $-3$ mass 
dimensions. The proton decay amplitude through the gauge-boson exchange 
is given by 
\begin{equation}
 d^4 x \; d^4 x' \; \; g_7^2 J_i^\mu(x) J_j^\nu(x') \; \eta_{\mu\nu}
  G(x-x'; \vec{y}_i, \vec{y}_j). 
\end{equation}
Here, $G(x-x'; \vec{y}_i, \vec{y}_j)$ is a Green function on 
$\R^{3,1} \times Q$, and is approximately 
\begin{equation}
 G(x-x'; \vec{y}_i, \vec{y}_j) \sim 
   \frac{1}{{\rm vol}(Q)} \sum_{\vec{q}} 
   \int \frac{d^4p}{(2\pi)^4} \frac{-i}{p^2 - |\vec{q}|^2 }
   e^{-i p \cdot (x-x')} e^{i \vec{q} \cdot (\vec{y}_i - \vec{y}_j)}.
\label{eq:GreenFcn-Q}
\end{equation}
$\vec{q}$ labels eigenmodes of Laplace operator on $Q$, and their 
eigenvalues (Kaluza--Klein masses-square) and eigenfunctions are 
denoted by 
$|\vec{q}|^2$ and $e^{i \vec{q} \cdot \vec{y}}/\sqrt{{\rm vol}(Q)}$, 
respectively. 
The momentum transfer $p^\mu$ in $\R^{3,1}$ direction is 
only of order $1 \; \GEV$ in proton decay process. 
Since $p^2$ in the propagator is much smaller than the Kaluza--Klein 
masses $|\vec{q}|^2$, $p^2$ can be ignored and dropped from the
propagator. Carrying out $d^4 p$ integration, one obtains 
dimension-6 operators in an effective theory:
\begin{equation}
 d^4 x \, J_i^\mu(x) J_{j \mu}(x) \; 
 \left[g_7^2 G_{int}(\vec{y}_i, \vec{y}_j) \right] \sim 
 d^4 x \, J_i^\mu(x) J_{j \mu}(x) \; 
 \left[g_7^2 \frac{1}{|\vec{y}_i - \vec{y}_j|} \right].  
\end{equation} 
$G_{int}$ is the Green function on the internal space $Q$.
The coefficient in the square bracket has mass dimension $-2$.

For two different currents $i \neq j$, the effective (mass)$^{-2}$ 
parameter is of order 
\begin{equation}
 \frac{g_7^2}{R} \sim \frac{g_7^2/R^3}{1/R^2} \sim 
 \frac{g_{\rm GUT}^2}{M_{\rm GUT}^2},
\end{equation}
where $|\vec{y}_i - \vec{y}_j|$ is set to a typical Kaluza--Klein radius 
of $Q$, $R$, and we assumed that the $\SU(5)_{\rm GUT}$ symmetry is broken by 
topological gauge field configuration on $Q$ (such as Wilson line), 
and hence $M_{\rm GUT} \sim M_{\rm KK} \sim 1/R$. Thus, there is 
no enhancement compared with typical proton decay amplitude through 
gauge-boson exchange in 4D field theory models. 
For the same current, $i = j$, however, $1/|\vec{y}_i - \vec{y}_i|$
diverges. The amplitude diverges (linearly), because all the 
Kaluza--Klein modes 
with arbitrary large momentum $\vec{q}$ equally contribute without 
cancellation in (\ref{eq:GreenFcn-Q}) in case $\vec{y}_i = \vec{y}_j$.
In reality, however, 
localized charged matter fields have certain form factor, or put 
differently, intersecting D6-branes effectively have certain 
``thickness''. The Kaluza--Klein momentum sum in (\ref{eq:GreenFcn-Q}) 
is effectively cut-off at around $|\vec{q}| \sim M_*$, where $M_*$ 
is the string scale, and the effective (mass)$^{-2}$ parameter becomes 
\begin{equation}
 g_7^2 M_* \sim \frac{g_7^2 / R^3}{1/R^2} (R M_*) \sim 
 \frac{g_{\rm GUT}^2}{M_{\rm GUT}^2} \times 
    \left(\frac{M_*}{M_{\rm GUT}}\right).
\end{equation}
The effective dimension-6 operator for the same current, 
${\bf 10}^\dagger {\bf 10} {\bf 10}^\dagger {\bf 10}$ in the effective 
K\"{a}hler potential, has a coefficient enhanced by $(M_*/M_{\rm KK})$ 
\cite{friedmanwitten}.

Let us now study the enhancement factor in F-theory compactifications. 
We begin with the ${\bf 10}^\dagger {\bf 10} {\bf 10}^\dagger {\bf 10}$ 
operator. In supersymmetric F-theory compactifications, chiral
multiplets in the $\SU(5)_{\rm GUT}$-{\bf 10} representation correspond 
to holomorphic sections $f_i$ of a line bundle on the matter curve 
$\bar{c}_{({\bf 10})}$ in a complex surface $S$ of $A_4$ singularity 
\cite{Curio, DI}. Here, $i= 1,2,3$ is now the generation index of chiral
multiplets in the $\SU(5)_{\rm GUT}$-{\bf 10} representation.
We take the coordinates on $S$ as $y_{1,2}$ and $w_{1,2}$, where 
$y_{1,2}$ correspond to normal directions of the matter curve 
$\bar{c}_{({\bf 10})}$, and $w_{1,2}$ to coordinates on the curve. 
The gauge fields on $S$ and the chiral zero modes in the {\bf 10}
representation couple as 
\begin{equation}
 \int_{\R^{3,1}} d^4 x \int_S d^2y d^2w \; \; 
    J^\mu_{ji}(x) \chi_i(y,w) \chi_j^*(y,w) A_\mu(x,y,w),
\end{equation}
where $J^\mu(x)_{ji}$ is a dimension-3 current 
$\bar{\lambda}_j \bar{\sigma}^\mu \lambda_i$ on $\R^{3,1}$, 
where $\lambda_i(x)$ and $\lambda_j(x)$ are fermions in the 
effective theory corresponding to the zero modes $f_i(w)$ 
and $f_j(w)$. $A_\mu(x,y,w)$ is the gauge field on $S$, and 
is assigned a mass-dimension 1. Its kinetic term is 
\begin{equation}
- \frac{1}{4 g_8^2} \int_{\R^{3,1}} d^4 x \int_S d^2 y d^2 w \; \; 
 \tr (F_{MN} F^{MN}).
\end{equation}
$\chi_{i,j}(y,w)$ are the zero-mode wavefunctions on $S$, corresponding 
to $f_{i,j}$ (see \cite{Hayashi-2}). Their approximate form, as well as 
their normalization, are 
\begin{equation}
 \chi_{i,j} \sim e^{- M_*^2 |y|^2} f_{i,j}(w) 
   \left(\frac{M_*}{\sqrt{{\rm vol}(\bar{c}_{({\bf 10})})}}\right).
\label{eq:wavefcn-10}
\end{equation}
Thus, the proton decay amplitude becomes 
\begin{eqnarray}
 & & 
   d^4x d^4 x' J^\mu_{ji}(x)J_{\mu lk}(x')  \nonumber \\
 & &   \quad   \int_S d^2y d^2 w\int_S d^2 y' d^2 w' 
   g_8^2 \chi_i(y,w) \chi^*_j(y,w) \chi_k(y',w') \chi_l^*(y',w')
   G(x-x'; y,y',w,w'),
\end{eqnarray}
and the approximate form of the Green function is 
\begin{equation}
 G(x-x'; y,y',w,w') \sim \frac{1}{{\rm vol}(S)} \sum_{\vec{k},\vec{q}} 
   \int \frac{d^4 p}{(2\pi)^4} \frac{-i}{p^2 - |\vec{k}|^2 - |\vec{q}|^2}
   e^{-ip \cdot (x-x')} e^{i \vec{q}\cdot (\vec{y}- \vec{y}')}
   e^{i \vec{k} \cdot (\vec{w} - \vec{w}')}.
\label{eq:GreenFcn-F}
\end{equation}
$p^2$ in the propagator is negligible (just like in the case of M-theory
compactifications), and the $d^4 p$ integration can be carried out. 
Thus, we obtain a dimension-6 operator 
\begin{equation}
 d^4 x \; \; J^\mu_{ji}(x) J_{\mu lk}(x) 
 \left[\int_S d^2y d^2w \int_S d^2y' d^2 w' \; \; g_8^2
   (\chi_j^* \chi_i)(y,w) (\chi_l^* \chi_k)(y',w') G_{int}(y,y',w,w')\right]. 
\label{eq:86}
\end{equation}

To evaluate the effective (mass)$^{-2}$ parameter in the square bracket, 
we proceed as follows. Because of the fact that 
the {\bf 10}-representation fields are localized along the curve, or 
equivalently because of the exponentially falling off wavefunctions 
$\chi_{i,j,k,l}$ in (\ref{eq:wavefcn-10}), dominant contribution 
to the amplitude comes from a region where $\vec{y}$ and $\vec{y}'$ 
are close to each other (and also to the matter curve where 
$\vec{y}\sim \vec{y}' \sim \vec{0}$), very large $\vec{q}$ can 
contribute in (\ref{eq:GreenFcn-F}). On the other hand, 
all the {\it zero} modes in ${\bf 10}$-representation are characterized by 
holomorphic (and hence smooth) sections $f_{i,j,k,l}(w)$ of a line
bundle (without a torsion component), and only low-lying Kaluza--Klein
momenta $\vec{k}$ can contribute after $d^2 w d^2 w'$
integration.\footnote{This is where our analysis is different from 
that of \cite{DW2}.}
Thus, we ignore $|\vec{k}|^2$ 
and keep only $|\vec{q}|^2$ in the denominator of (\ref{eq:GreenFcn-F}). 
Now, summation in $\vec{k}$ can be carried out, and $G_{int}$
becomes proportional to $\delta^2(\vec{w} - \vec{w}')$. 
The Kaluza--Klein momentum sum in $\vec{q}$ yields a logarithmic divergence, 
which is cut-off at $|\vec{q}| \sim M_*$ because of the thickness of 
the Gaussian wavefunction in (\ref{eq:wavefcn-10}).
In the end, the amplitude looks 
\begin{equation}
 d^4 x J^\mu_{ji}(x) J_{\mu lk}(x) 
   \frac{g_8^2}{({\rm vol}(\bar{c}_{({\bf 10})}))^2} \int d^2 w 
    ( f_j^* f_i f_l^* f_k )(w) \times 
   \ln \left(\frac{M_*^2}{M_{\rm KK}^2}\right).
\end{equation}
Since the factor 
\begin{equation}
 \frac{g_8^2}{{\rm vol}(\bar{c}_{({\bf 10})})} \sim 
 \frac{g_8^2/R^4}{1/R^2} \sim \frac{g_{\rm GUT}^2}{M_{\rm GUT}^2}
\end{equation}
is the usual effective (mass)$^{-2}$ scale of the dimension-6 
proton decay operator, the enhancement factor in F-theory 
compactifications is $\ln (M_*/M_{\rm GUT})^2$. 
This logarithmic enhancement factor\footnote{
If one ignores the difference between $M_*$, $1/l_s$ or
$1/\sqrt{\alpha'}$, and sets $g_s \sim 1$, then 
$(M_*/M_{\rm GUT})^4 \sim 1/\alpha_{\rm GUT} \sim 24$. 
Thus, the enhancement factor in the amplitude is of order 
$\ln (1/\sqrt{\alpha_{\rm GUT}}) \sim \ln 5$. 
This result differes from the $(1/\alpha_{\rm GUT})^{1/2}$ 
enhancement (which corresponds to quadratic divergence) in \cite{DW2}. 
Quantitatively, 
$\ln \sqrt{1/\alpha_{\rm GUT}} \sim \ln 5$ is about a factor 4 smaller 
than $1/\sqrt{\alpha_{\rm GUT}} \sim 5$ in the decay amplitude, and the 
decay rate based on logarithmic enhancement is about an order of
magnitude smaller than that based on quardratic enhancement.} 
originates 
from the fact that chiral multiplets in the 
$\SU(5)_{\rm GUT}$-{\bf 10} representation are localized relatively 
to the $\SU(5)_{\rm GUT}$ gauge fields in real two dimensions.

Finally, we study the enhancement factor associated with the 
effective dimension-6 proton decay operator 
${\bf 10}^\dagger_j {\bf 10}_i \bar{\bf 5}^\dagger_b \bar{\bf 5}_a$.
Here, $a,b$ are generation indices.
Chiral multiplets in the $\SU(5)_{\rm GUT}$-$\bar{\bf 5}$ representation 
are also described by holomorphic sections $h_{a,b}$ of 
a {\it line bundle} on a curve
$\tilde{\bar{c}}_{(\bar{\bf 5})}$, which is obtained by resolving 
all the double point singularities of the matter curve 
$\bar{c}_{(\bar{\bf 5})}$ at the codimension-3 loci of enhanced $D_6$ 
singularity \cite{Hayashi} (called type (d) points there). 
In light of Heterotic--F theory duality, $\bar{\bf 5}$'s may well be 
described only as sections of a {\it sheaf} on $\bar{c}_{(\bar{\bf 5})}$, 
and the sheaf may not be torsion free or locally free, in principle. 
Whether such a localized component exists in the sheaf on the curve
$\bar{c}_{(\bar{\bf 5})}$, and hence in the zero modes, is crucial 
for the analysis of the enhancement factor in proton decay amplitude. 
Reference \cite{Hayashi} concluded, however, that there is not a 
localized component at all; all the zero modes are described by {\it smooth} 
sections $h_{a,b}$ on the covering matter curve 
$\tilde{\bar{c}}_{(\bar{\bf 5})}$.

The current of zero modes of $\bar{\bf 5}$'s couple to 
the $\SU(5)_{\rm GUT}$ gauge field through 
\begin{equation}
 \int_{\R^{3,1}} d^4 x \int_S d^2 y d^2 w \; J^\mu_{ba}(x) 
  \chi_b^*(y,w) \chi_a(y,w) A_\mu(x,y,w).
\end{equation}
Here, $J^\mu_{ba}$ is a current on $\R^{3,1}$ that consists of 
fermions $\lambda_{a,b}$ corresponding to the zero modes $h_{a,b}$, 
and $\chi_{a,b}$ are their zero-mode wavefunctions (see
\cite{Hayashi-2}). The dominant contribution to the 
${\bf 10}^\dagger {\bf 10} \bar{\bf 5}^\dagger \bar{\bf 5}$ decay 
amplitude most likely comes from a region around intersection points 
of the two matter curves, $\bar{c}_{({\bf 10})}$ and 
$\bar{c}_{(\bar{\bf 5})}$. Although there are two different types of 
intersection points (type (a) and type (d) points in the classfication
of \cite{Hayashi}), the difference will not matter for the proton
decay that takes place within $\SU(5)_{\rm GUT}$. Thus, here, we assume 
that the matter curve $\bar{c}_{({\bf 10})}$ is along $y = 0$ (locally), 
and $\bar{c}_{(\bar{\bf 5})}$ along $w = 0$. Then, the wavefunction 
$\chi_{a,b}$ becomes approximately 
\begin{equation}
 \chi_{a,b}(y,w) \sim e^{- M_*^2 |w|^2} h_{a,b}(y) 
    \times \left(\frac{M_*}{R}\right).
\end{equation}

Repeating the same process as before, one finds that the 
effective dimension-6 operator is given by 
\begin{eqnarray}
& & d^4 x \; J^\mu_{ji}(x) J_{\mu ba}(x) \; 
 \int_S d^2 y  d^2 w \int_S d^2 y' d^2 w' \; \; 
 g_8^2  \nonumber \\
 & & \quad \frac{1}{R^4} \sum_{\vec{k}, \vec{q}} 
   \left(\frac{M_*}{R}\right)^4 
   (f_j^* f_i)(w) e^{-M_*^2 |y|^2}
   (h_b^* h_a)(y') e^{- M_*^2 |w'|^{2}} 
    \frac{i}{|\vec{k}|^2 + |\vec{q}|^2}
    e^{i \vec{k}\cdot (\vec{w}-\vec{w}')} 
    e^{i \vec{q} \cdot (\vec{y}- \vec{y}')}.
\label{eq:91}
\end{eqnarray}
Since $h_{a,b}(y')$ and $f_{i,j}(w)$ are zero modes, 
and are {\it smooth} everywhere along the matter curves, 
only Kaluza--Klein gauge bosons with low-lying Kaluza--Klein 
momenta $\vec{q}$ AND $\vec{k}$ couple to both $\bar{\bf 5}$'s 
and ${\bf 10}$'s. [Remember that we have $d^2y' d^2 w$
integration in the expression above.]\footnote{This is where our study
differs from that in \cite{DW2}.}
Thus, the infinite sum in $\vec{q}$ and 
$\vec{k}$ is effectively dropped, $\vec{q}$ and $\vec{k}$ 
replaced by $1/R$, and we find that the effective coefficient 
of the dimension-6 operator is of order 
\begin{equation}
 \frac{g_8^2}{R^2} \sim \frac{g_8^2/R^4}{1/R^2} \sim 
 \frac{g_{\rm GUT}^2}{M_{\rm GUT}^2}.
\end{equation}
Therefore, the prediction of the 
${\bf 10}^\dagger_j {\bf 10}_i \bar{\bf 5}^\dagger_b \bar{\bf 5}_a$ 
dimension-6 proton decay in F-theory compactifications is just as 
the same as that of the ordinary GUT dimension-6 proton decay. 
There is no particular enhancement factor for this mode;\footnote{
This conclusion differs from the result, $1/\sqrt{\alpha_{\rm GUT}}$ 
enhancement, in \cite{DW2}.} this is 
essentially because only low-lying Kaluza--Klein modes can couple 
to both zero modes in the ${\bf 10}$ representation and those 
in $\bar{\bf 5}$. 

To conclude, 
$\Delta K = {\bf 10}_j^\dagger {\bf 10}_i {\bf 10}^\dagger_l {\bf 10}_k$ 
dimension-6 proton decay amplitude has a logarithmic enhancement. The 
enhancement factor is $\ln (M_*/M_{\rm KK})^2$, which is roughly 
$\ln 5 \sim 1.6$, where we used $1/\alpha_{\rm GUT} \sim 25$, and 
ignored a difference among $M_*$, $1/l_s$ and $1/\sqrt{\alpha'}$.
On the other hand, 
$\Delta K = {\bf 10}^\dagger_j {\bf 10}_i \bar{{\bf 5}}^\dagger_b
\bar{\bf 5}_a$ amplitude is dominated by low-lying Kaluza--Klein gauge
bosons, and is not enhanced. Thus, the decay rates to left-handed
positively charged leptons $(\ell^+)_L$ in ${\bf 10}$ are enhanced
typically by a factor of $1.6^2 \sim (2\sim 3)$, relatively to rates 
of decay to right-handed positively charged leptons, $(\ell^+)_R$, or 
to right-handed anti-neutrinos, $\bar{\nu}_R$ in $\bar{\bf 5}^\dagger$
(c.f. \cite{friedmanwitten}). It should be noted, however, that 
the decay amplitudes have generation-dependent factors\footnote{
These expressions should not be taken literally. It should be reminded 
that $f_k$ and $f_l$ for the current $J^\nu_{lk}$ correspond to 
zero modes in different irreducible representations of the Standard 
Model gauge group, although they are in the same irreducible
representation ${\bf 10}$ under $\SU(5)_{\rm GUT}$. Thus, $f_k$ is not
necessarily the same as $f_l$ even when $k = l$. The same is true 
for the zero modes $h_a$ and $h_b^*$ in the current $J^\nu_{ba}$ for
fermions in the $\SU(5)_{\rm GUT}$-$\bar{\bf 5}$ representation. It
should also be clear from the discussion in the main text that 
$e^{i \vec{k} \cdot \vec{w}}$ and $e^{i \vec{q} \cdot \vec{y}'}$ with 
low-lying Kaluza--Klein momenta $\vec{k}$ and $\vec{q}$ are omitted from
the second factor. } 
\begin{equation}
 \frac{1}{{\rm vol}(\bar{c}_{({\bf 10})})} \int d^2 w 
     (f^*_j f_i f^*_l f_k)(w) \quad {\rm and} \quad 
 \frac{1}{{\rm vol}(\bar{c}_{({\bf 10})})} \int d^2 w 
     (f^*_j f_i)(w) \; 
 \frac{1}{{\rm vol}(\bar{c}_{(\bar{\bf 5})})} \int d^2 y'
     (h^*_b h_a)(y') 
\label{eq:unknown}
\end{equation}
for the ${\bf 10}^\dagger_j {\bf 10}_i {\bf 10}^\dagger_l {\bf 10}_k$
and ${\bf 10}^\dagger_j {\bf 10}_i \bar{\bf 5}^\dagger_b \bar{\bf 5}_a$ 
processes, respectively, and these factors may well be more important 
for individual decay modes than the logarithmic enhancement factor. 
Thus, the logarithmic enhancement of decay rates to charged leptons 
should be regarded only as a tendency predicted among all the decay modes. 

As for the total decay rate of proton through the gauge-boson exchange,
the enhancement remains only logarithmic, and is of order a factor of 2--3.
It is not even clear whether this enhancement is more important than the 
yet to be (and hard to be) calculated factors in (\ref{eq:unknown}), which 
may result in suppression. More important is a fact that the total decay 
rate is proportional to $M_{\rm GUT}^4$, and that the value of $M_{\rm GUT}$
still has a large uncertainty, ranging from, say, $10^{15.7} \; \GEV$ 
to $10^{16.5} \; \GEV$ (see Figure~\ref{fig:unif}).
The decay rate for 
$M_{\rm GUT} = 10^{15.75} \; \GEV$ is three orders of magnitude larger 
than that for 
$M_{\rm GUT} = 10^{16.5} \; \GEV \simeq 3 \times 10^{16} \; \GEV$. 
In the scenario of $\SU(5)_{\rm GUT}$ symmetry breaking discussed in
this article, $M_{\rm GUT}$ tends to be small, because of the tree-level 
correction to the gauge couplings, and hence the decay rate tends to 
be large. Such model-dependence is more important in the total decay
rate than the logarithmic enhancement that is applied to all the
F-theory models of $\SU(5)_{\rm GUT}$. Thus, the total decay rate 
can be used in discriminating various models, and the ratio of rates 
of decays to charged leptons to rates of decays to anti-neutrinos 
can be used to see whether charged matter fields are localized 
in internal space dimensions or not.


\section*{Acknowledgements} 

We thank Aspen Center for Physics for hospitality, where 
this project started during a summer programme in 2006. 
This work is supported in part by PPARC (RT), 
by the US DOE under contract No. DE-FG03-92ER40701 and  
the Gordon and Betty Moore Foundation (till 12, 2006), 
and by the World Premier International Research Center Initiative 
(WPI Initiative), MEXT, Japan (TW).

\appendix

\section{Interpretation of ``Wilson Lines'' in Toroidal Orbifolds}

Model building using toroidal orbifold has a long history that 
dates back to 1980's. Geometry of orbifolds is understood as certain 
limits of Calabi--Yau manifold. It has been known since early 
days (e.g. \cite{WittenOrbifold, Vafa}) that toroidal orbifold
compactifications of the Heterotic theory corresponds to 
some limits in the moduli space of compactifications given by 
a Calabi--Yau manifold and a vector bundle on it.
Toroidal orbifolds in the context of Type IIB orientifold with D7-branes 
and O7-planes are little more involved in its interpretation as limits of 
smooth Calabi--Yau orientifold,\footnote{Some of fractional D3-branes 
are interpreted as anti D7-brenes with a vector bundle on them, and 
such fractional D3-branes do not remain stable when vanishing cycle
is blown up to be large.} yet some works have already been done. 

In the appendix of this article, we clarify how one should interpret 
``Wilson lines'' in toroidal orbifold compactifications of the Heterotic
string theory in terminology of smooth Calabi--Yau compactifications.
Toroidal orbifold models using discrete Wilson lines gained a renewed 
attention triggered by an activity that followed papers on 
$S^1/\Z_2 \times \Z'_2$ orbifold GUT \cite{Kawamura, HN, HMN}.
We will see in Heterotic theory compactification that the toroidal 
orbifolds with ``discrete Wilson lines'' are also understood as some 
limits of compactifications described by smooth Calabi--Yau manifold 
and a vector bundle on it.
The discrete Wilson lines in toroidal orbifolds are not 
Wilson lines (or flat bundles) on smooth Calabi--Yau $Z$ 
associated with a discrete homotopy group $\pi_1(Z)$, but rather 
they correspond to turning on line bundles on a Calabi--Yau with 
the U(1) structure group of the line bundles chosen differently at 
different vanishing cycles buried at orbifold singularities.\footnote{
See e.g. \cite{Recent} for recent studies on this subject.}  

Once one adopts the interpretation above, then the $\SU(5)_{\rm GUT}$ 
symmetry breaking in Heterotic toroidal orbifold compactifications 
(with or without discrete Wilson lines) are regarded as special cases 
of the material discussed in the main text. Thus, as we discuss 
in the appendix~\ref{sssec:discuss} (and as one can understand 
as special cases of the discussion in section~\ref{ssec:Y-problem}), 
so-called the toroidal orbifold GUT's in Heterotic string theory 
also suffer from the $\U(1)_Y$ problem.

In the literature of toroidal orbifolds, another terminology 
``continuous Wilson line'' is also found. Although the continuous 
Wilson lines have nothing to do with the main theme of this article, 
we take this opportunity (in the appendix \ref{ssec:continuous}) 
to clarify that the ``continuous Wilson lines'' 
in Heterotic toroidal orbifold correspond to a part of vector bundle moduli 
in smooth Calabi--Yau compactification.

\subsection{Discrete Wilson Lines}
\label{ssec:discrete}

Since our motivation is to understand what the ``discrete Wilson lines''
really are, we do not have to work on a very realistic model. 
Simple examples that illustrate the point will be better suited for 
our purpose. Thus, we use $T^4/\Z_k$ orbifolds instead of $T^6/\Z_N$ 
orbifolds, and provide interpretations of discrete Wilson lines 
in terms of compactification on K3 surfaces with vector bundles on them.
K3 compactification [$T^4/\Z_k$ in orbifold limits] has an advantage over 
$CY_3$ [resp. $T^6/\Z_N$] compactification in that index theorem can 
calculate the massless spectrum of vector bundle moduli in addition 
to that of charged multiplets, so that we can compare the number of 
vector bundle moduli of smooth manifolds with that of orbifolds. 
We also use the Heterotic $\SO(32)$ string theory, instead of 
$E_8 \times E'_8$, because we are not trying to analyse geometry of 
specific toroidal orbifolds to be used for semi-realistic models, 
but we try to understand what the discrete Wilson lines are. 
For that purpose, difference in the choice of gauge group is not 
a big deal.
We calculate the massless spectrum both in K3+bundle compactification 
and in toroidal orbifolds and confirm that the results do agree. 
The agreement shows that the K3+bundle interpretation is correct
for the toroidal orbifolds of the Heterotic theory, 
and at the same time tells us the geometric meaning of twisted sector 
fields. 

\subsubsection{Spectrum of Smooth-Manifold Compactification}
\label{sssec:spec-field}

Let us consider a Heterotic SO(32) string theory compactified on 
a K3 manifold $Z$, with a vector bundle $V$ turned on.
The D = 10 supergravity multiplet reduces to 
\begin{itemize}
\item D = 6 supergravity multiplet and a D = 6 tensor multiplet, 
containing D = 6 metric, one 2-form field and one scalar.
\item $h^{1,1}(Z)=20$ hypermultiplets, containing $3 \times 19$ real scalars 
describing the deformation of the metric of $Z$, 22 scalars obtained 
by integrating $B$-field over the 22 2-cycles of $Z$, and one more
      scalar \cite{SeibergAspinwall}. 
\end{itemize}
When the structure group of the vector bundle $V$ is 
$\SO(2r) \subset \SO(32)$, $\SO(32-2r)$ is the unbroken symmetry, and 
the SO(32)-adjoint representation decomposes into 
\begin{equation}
 \mathfrak{so}(32)\mbox{-}{\bf adj.} \rightarrow 
 ({\bf 1},\mathfrak{so}(32-2r)\mbox{-}{\bf adj.}) +
 (\mathfrak{so}(2r)\mbox{-}{\bf adj.},{\bf 1}) + 
 ({\bf vect.},{\bf vect.}).
\end{equation}
The multiplicity of hypermultiplets is calculated by indices 
\begin{equation}
 - \frac{1}{2}\int_Z {\rm ch}_R (V) \hat{A}(TZ) = 
 T_R I_V - ({\rm dim.}R) \int_Z \frac{c_2(TZ)}{24}
= 24 \, T_R - ({\rm dim.}R),
\end{equation}
where $T_R$ is a Dynkin index\footnote{$T_R$ is 1 for vector representations 
and $2r-2$ for adjoint representations of $\SO(2r)$, and 
$1/2$ for fundamental representations and $N$ for adjoint representations 
of $\SU(N)$.}, $I_V \equiv - (2 T_R)^{-1} \int_Z {\rm ch}_{2,R}(V)$ 
is the instanton number of the bundle $V$, and 
$I_V = \int_Z c_2(TZ) = 24$ is used at the last equality.
The D = 10 SO(32) vector multiplet reduces to 
\begin{itemize}
\item one D = 6 $\SO(32-2r)$ vector multiplet
\item $(24-2r)$ hypermultiplets of $\SO(32-2r)$-vector representation, 
\item $24(2r-2) - r(2r-1)$ hypermultiplets of vector bundle moduli.
\end{itemize}

Let us check the Higgs cascade, as in the analysis of \cite{KV,6authors}. 
As one of hypermultiplets in the vector representation develops 
an expectation value, the unbroken symmetry becomes $\SO(32-2(r+1))$.
Each hypermultiplet in the $[32-2r]$-dimensional vector representation 
reduces into a hypermultiplet in the $[32-2(r+1)]$-dimensional vector 
representation of $\SO(32-2(r+1))$ unbroken symmetry and 2 singlets. 
The Higgs mechanism associated with the symmetry breaking 
$\SO(32-2r) \rightarrow \SO(32-2(r+1))$ absorbs 2
hypermultiplets in the $\SO(32-2(r+1))$-vector representation and 
one singlet. 
Thus, $[(24-2r)-2] =[24-2(r+1)]$ hypermultiplets in the 
vector representation are left after the symmetry breaking, 
which agrees with the result of $\SO(2(r+1))$-bundle compactification.
Similarly, the number of singlet hypermultiplet---vector bundle 
moduli---increases by $2 \times (24 - 2r) - 1$, because 
two singlets arise from one hypermultiplet in the vector representation, 
but one hypermultiplet is absorbed by a Higgsed vector multiplet.
Thus, the number of vector bundle moduli hypermultiplets becomes 
$[24(2(r+1)-2)-(r+1)(2(r+1)-1)]$ after the symmetry breaking 
$\SO(32 - 2r) \rightarrow \SO(32 - 2(r+1))$, which also agrees 
with the result of $\SO(2(r+1))$ bundle compactification.
Moduli spaces of different unbroken symmetry and different structure 
group are continuously connected through this Higgs cascade process. 

Case with $r=2$, however, needs a separate treatment, because the 
structure group of a rank-4 vector bundle $\SO(4) \simeq 
\SU(2) \times \SU(2)$ is not a simple group.
The rank-4 bundle is a tensor product $V \simeq V_1 \otimes V_2$, 
and the instanton number is given by 
\begin{equation}
 I_V = I_{V_1} + I_{V_2}.
\end{equation} 
One can see that the numbers of $\SO(28)$-vector and singlet 
hypermultiplets given above are correct also for the $r=2$ cases, 
if $I_{V_1}$ and $I_{V_2}$ are both non-zero.
If the instanton number is only in either one of SU(2), say, 
$I_{V_2} = 0$, however, the unbroken symmetry group is 
$\SU(2) \times \SO(28)$, and there are 
$[T_{V_1} I_{V_1} - {\rm dim.} V_1] = (24-4)/2=10$ hypermultiplets 
in the $({\bf 2},{\bf 28})$ representation and $2 \times 24 - 3 = 45$ 
vector bundle moduli.

\subsubsection{Spectrum of Orbifold Compactification}
\label{sssec:spec-orbifold}

Let us now calculate massless spectra of some of $T^4/\Z_k$ orbifolds, 
and compare them with what we have got from the field-theory calculation.
The Heterotic SO(32) string theory is described by bosons on the 
worldsheet, $X^\mu$ ($\mu = 0,1,2,3$), $Z^A$, $\overline{Z}^{\bar{A}}$ 
($A=1,2$), right-moving fermions, $\psi^{\mu}$, $\psi^A$, 
$\overline{\psi}^{\bar{A}}$, and left-moving fermions, $\lambda^I$, 
$\overline{\lambda}^{I}$ ($I=1,\cdots,16$).
Toroidal orbifolds $T^4/\Z_k$ ($k=2,3,4,6$) are quotients 
$\C^2/(\Z_k\vev{\sigma} \ltimes \Lambda)$, where 
$\Lambda$ is a rank-4 lattice in $\C^2$ whose basis consists of 4 vectors 
$e^A_a$ ($a = 1,2,3,4$) and $\sigma$ is an $\SU(2) \subset \SO(4)$ rotation 
on $\C^2$, satisfying $\sigma^k = {\bf id.}$.
The worldsheet fields $Z^A$ and $\overline{Z}^{\bar{A}}$ transform under 
the generators of the space group $\Z_k \ltimes \Lambda$ as 
\begin{eqnarray}
 \tau_a : Z^A \rightarrow Z^A + e^A_a, & \quad & 
 \sigma : Z^A \rightarrow e^{2\pi \, i v^A} Z^A, \\ 
 \tau_a : \overline{Z}^{\bar{A}} \rightarrow \overline{Z}^{\bar{A}} 
    + e^{\bar{A}}_a, & \quad &
 \sigma : \overline{Z}^{\bar{A}} \rightarrow e^{-2\pi \, i v^A} 
      \overline{Z}^{\bar{A}},
\end{eqnarray}
where $\tau_a$ ($a=1,2,3,4$) are translation along the vectors $e_a$, 
and $e^{\bar{A}}_a$ are complex conjugates of $e^A_a$.
$\sigma$ is a generator of rotation on the complex coordinates, 
and $v^A = (1/k,-1/k)$. 
Other fields on the worldsheet transform under the translation and 
rotation as 
\begin{eqnarray}
 \tau_a : \psi^A \rightarrow \psi^A, \qquad & \quad & 
 \sigma : \psi^A \rightarrow e^{2\pi \, i v^A} \psi^A, \\
 \tau_a : \overline{\psi}^{\bar{A}} \rightarrow \overline{\psi}^{\bar{A}}, 
     \qquad & \quad & 
  \sigma : \overline{\psi}^{\bar{A}} \rightarrow e^{-2\pi \, i v^A} 
     \overline{\psi}^{\bar{A}},\\
 \tau_a : \lambda^I \rightarrow e^{2\pi \, i W^I_a} \lambda^I, 
     & \quad & 
 \sigma : \lambda^I \rightarrow e^{2\pi \, i V^I} \lambda^I, \\
 \tau_a : \overline{\lambda}^I \rightarrow e^{-2\pi \, i W^I_a} 
    \overline{\lambda}^I,  & \quad & 
 \sigma : \overline{\lambda}^I \rightarrow e^{-2\pi \, i V^I} 
    \overline{\lambda}^I;     
\end{eqnarray}
all of $\beta_a \equiv \diag(e^{2\pi \, i W^I_a},e^{-2\pi \, i W^I_a})$ 
($a=1,2,3,4$) and 
$\gamma_\sigma \equiv \diag(e^{2\pi \, i V^I},e^{-2\pi \, i V^I})$ 
in SO(32) acting on $(\lambda^I,\overline{\lambda}^I)$ commute each other; 
although they do not have to commute as long as those matrices satisfy 
the algebra of $\tau_a$ and $\sigma$ in the space group, 
we only consider the simplest cases here.\footnote{Slightly more complicated 
examples---non-diagonal $\gamma_\sigma$---will be discussed in 
the appendix~\ref{ssec:continuous}.}
When $\diag(W^I_a,-W^I_a) \neq 0$, 
$W^I_a$ are called discrete Wilson lines. In toroidal compactification, 
$2 \pi W^I_a = A^I_A e^A_a + A^I_{\bar{A}} e^{\bar{A}}_a$ are the 
Wilson lines along the four independent topological 1-cycles of $T^4$. 
But, in (the blow up of) toroidal orbifolds $T^4/\Z_k$ ($k=2,3,4,6$), 
there are no topological 1-cycles.\footnote{The Euler number of 
a simply connected K3 manifold is 24, and the Euler number of the 
resolved $T^4/\Z_k$ should be $24/\# \pi_1(T^4/\Z_k)$, where 
the resolution of $T^4/\Z_k$ were to have a non-trivial homotopy group. 
The Euler number of the blow up of $T^4/\Z_k$ can be calculated 
(See e.g., \cite{GSW} for how to calculate the Euler number of 
toroidal orbifolds.) and is known to be 24 for all of $k=2,3,4,6$. 
Thus, all the $T^4/\Z_k$'s have trivial homotopy groups.
It is also possible to confirm that they 
are simply connected, by explicitly looking at the geometry of 
$A_{k-1}$-type ALE space expressed as $S^1$-fibration over a real 
three-dimensional space.} 
Thus, there is no way the ``Wilson lines'' have anything to do 
with a flat bundle associated with a non-trivial homotopy group.
The ``Wilson lines'' $W^I_a$ in toroidal orbifolds are allowed 
to take only discrete values, because of the algebraic relation 
between $\sigma$ and $\tau_a$ (and of the relation bewteen 
$\gamma_\sigma$ and $\beta_a$), and hence they are called 
``discrete Wilson lines'' in the literature of toroidal orbifold 
compactifications, but they are not Wilson lines associated with 
a non-trivial homotopy group $\pi_1(Z)$ that is allowed to take 
discrete values because of the discreteness of $\omega_1(Z)$.

{\bf Cases Without Discrete Wilson Lines}

Now that the notation is set, let us compute the massless spectrum 
of toroidal orbifolds. We discuss only $T^4/\Z_2$ and $T^4/\Z_3$
orbifolds for simplicity.
As a warming up, we start with cases without discrete Wilson lines. 
 toroidal orbifolds with the discrete Wilson linse are discussed later. 

One has to choose\footnote{We choose $0\leq V^I \leq 1$ for $I = 1,\cdots,16$.}
$\gamma_\sigma = 
\diag(e^{2\pi \, V^I},e^{- 2\pi \, i V^I})$ so that 
\begin{equation}
 \frac{1}{2} \left[ \sum_A |v^A|(1-|v^A|) - \sum_I V^I (1-V^I)\right] \equiv 0 
\qquad \left({\rm mod~} \frac{1}{k}\Z \right)
\label{eq:Vafa-orbifold}
\end{equation}
for a consistency on the spectrum of a $\sigma$-twisted sector \cite{Vafa}. 
For the $T^4/\Z_2$ orbifold ($k=2$), solutions are 
\begin{eqnarray}
 V^I_{r=2} & = & \frac{1}{2}\,(1,1,\overbrace{0,\cdots,0}^{14}), 
    \label{eq:Z2-r2-V}\\
 V^I_{r=6} & = & 
    \frac{1}{2}\,(\overbrace{1,\cdots,1}^6,\overbrace{0,\cdots,0}^{10}).
\end{eqnarray}
The spectrum is summarized as follows:
\begin{itemize}
\item $r=2$
 \begin{itemize}
 \item Untwisted sector
   \begin{itemize}
    \item D = 6 sugra and tensor multiplets,
    \item 4 hypermultiplets from D = 10 metric and $B$-field,
    \item $\SU(2) \times \SU(2) \times \SO(28)$ vector multiplet, 
    \item ({\bf 2},{\bf 2},{\bf 28}) hypermultiplet.
   \end{itemize}
  \item Twisted sector $\times 16$
   \begin{itemize}
     \item ({\bf 1},{\bf 2},{\bf 1}) 4 half hypermultiplets,
     \item ({\bf 2},{\bf 1},{\bf 28}) half hypermultiplet.
   \end{itemize}
 \end{itemize}
\item $r=6$
  \begin{itemize}
   \item Untwisted sector
    \begin{itemize}
    \item D = 6 sugra and tensor multiplets, 
    \item 4 hypermultiplets from D = 10 metric and $B$-field
    \item $\SO(12) \times \SO(20)$ vector multiplet, 
    \item ({\bf 12},{\bf 20}) hypermultiplet.
    \end{itemize}
   \item Twisted sector $\times 16$
     \begin{itemize}
       \item ({\bf spin},{\bf 1}) half hypermultiplet.
     \end{itemize}
  \end{itemize}
\end{itemize}

In the $r=2$ case, the symmetry is broken down to $\SU(2) \times \SO(28)$, 
if the $({\bf 1},{\bf 2},{\bf 1})$ half hypermultiplets develop expectation 
values. There are $(1/2)\times 16$ hypermultiplets in the 
$({\bf 2},{\bf 28})$ representation, and 
there are 2 from the untwisted sectors; there are 10 hypermultiplets
in the $({\bf 2}, {\bf 28})$ representation as a whole in the toroidal 
orbifold calculation. 
This agrees with the field theory prediction in 
section~\ref{sssec:spec-field} for the case with instantons 
contained only in one of $\SU(2)$'s in 
$\SO(2r=4) \simeq \SU(2) \times \SU(2)$.\footnote{
$V^I_{r=2}$ corresponds to the embedding of the spin connection, and 
hence the instanton number is in only one of SU(2), not distributed 
in both SU(2)'s.}.

The twisted sectors and untwisted sector contribute to 
$\SU(2) \times \SO(28)$-singlet moduli hypermultiplets by $4 \times 16-3$ and 
$4$, respectively, and there are 65 as a whole. They correspond to 
$3 \times 16 - 3 = 45$ vector bundle moduli and $16 + 4=20 =
h^{1,1}(K3)$ bulk moduli, as we obtained in section
\ref{sssec:spec-field} for the case with an $\SU(2)$ bundle. 
Roughly speaking, each twisted sector has 4 moduli hypermultiplets, 
and one of them describes the blow up of the 
$\C^2/\Z_2$ singularity\footnote{\label{fn:blowup}
$k-1$ 2-cycles are burried in 
a $\C^2/\Z_{k}$ singularity. For each 2-cycle, there are three 
degrees of freedom of deforming metric, and value of 
$B$-field integrated over the 2-cycle is another freedom. Those four 
scalar degrees of freedom in $D=6$ effective theory form one 
hypermultiplet for such a 2-cycle. (See e.g., \cite{DM}.) 
Thus, there is one hypermultiplet corresonding to a $\C^2/\Z_2$ 
singularity.}. Thus, the remaining three twisted sector hypermultiplets 
(at each $\C^2/\Z_2$ fixed point) describe deformation of the 
vector bundle.

In the $r=6$ case, expectation values can be given to the
hypermultiplets in the SO(12)-{\bf spin} representation, so that the 
the SO(12) symmetry is completely Higgsed. Let us compare the massless 
spectra of toroidal orbifold and field-theory prediction in such a
situlation, only the $\SO(20)$ gauge symmetry is left unbroken.
The SO(20)-{\bf vect.} hypermultiplets arise only from the untwisted sector, 
and there are 12 as a whole, once again in agreement with the field-theory 
result, $24-2r=12$, in section \ref{sssec:spec-field}. The twisted and 
untwisted sectors yield $16 \times 16 - 66$ and $4$ moduli multiplets, 
and there are $190+4=194$ moduli in toroidal orbifold calculation.
This number of moduli is equal to the number of vector bundle moduli, 
$24 \times 10 - 66 = 174$, and the number of K3-moduli $16 + 4=20$ 
combined.

Thus, the number of moduli and the multiplicity of SO($32-2r$)-{\bf vect.} 
hypermultiplets are calculated both by field theory and by orbifold, and 
they agree. In the two examples of $T^4/\Z_2$ orbifolds we studied, 
the geometry $T^4/\Z_2$ is regarded as a particular limit of 
a K3 manifold, where 16 2-cycles are collapsed.
An example with $V^I_{r=6}$ is obtained by taking a limit further 
in the moduli space of $\SO(2r = 12)$ vector bundle on the K3-manifold, 
a limit where the structure group is reduced from $\SO(12)$ until the 
$\SO(12)$ symmetry is enhaced. 
Likewise, the toroidal orbifold compactification with $V^I_{r=2}$ can 
be approached from a field theory compactification, by taking a limit 
in the moduli space of $\SU(2) \subset \SO(4)$ vector bundle. It is 
a limit where the structure group is reduced from $\SU(2)$ until 
the $\SU(2)$ symmetry is enhanced and restored.

Let us also see examples of $T^4/\Z_3$ orbifolds. 
For the $T^4/\Z_3$ orbifold ($k=3$), solutions to the consistency condition 
(\ref{eq:Vafa-orbifold}) are
\begin{eqnarray}
V^I_{r=2} & = & \frac{1}{3}\,(1,1,\overbrace{0,\cdots,0}^{14}), 
  \label{eq:noWil-k3-r2} \\
V^I_{r=5} & = & 
   \frac{1}{3} \, (\overbrace{1,\cdots,1}^5,\overbrace{0,\cdots,0}^{11}),\\
V^I_{r=8} & = & 
   \frac{1}{3} \, (\overbrace{1,\cdots,1}^8,\overbrace{0,\cdots,0}^8).
  \label{eq:noWil-k3-r8}
\end{eqnarray}
The massless spectra of those models are:
\begin{itemize}
 \item Untwisted sector of $V^I_{r=2,5,8}$ models
   \begin{itemize}
    \item D = 6 sugra and tensor multiplets, 
    \item 2 hypermultiplets from D = 10 metric and $B$-field,
    \item $\SU(r) \times \SO(32-2r) \, \left[ \times U(1) \right]$ 
      vector multiplet,
    \item ({\bf r},{\bf vect.})$^1$ + ($\wedge^2$ {\bf r},{\bf 1})$^2$ 
      hypermultiplets,
   \end{itemize}
 \item Twisted sectors $\times 9$
   \begin{itemize}
    \item $r=2$: ({\bf 2},{\bf 28})$^1$ + $2 \times $ ({\bf 1},{\bf 1})$^2$ 
     + $5 \times $ ({\bf 1},{\bf 1})$^0$ hypermultiplets,
    \item $r=5$: ({\bf 1},{\bf vect.}) + $2 \times $ ({\bf 5},{\bf 1})$^1$ 
     + ($\wedge^2$ {\bf 5},{\bf 1})$^{-3}$ hypermultiplets,
   \item $r=8$: ($\wedge^2$ {\bf 8},{\bf 1})$^{-2}$ + 
     $2 \times $ ({\bf 1},{\bf 1})$^0$ hypermultiplets.
   \end{itemize}
\end{itemize}
By turning on expectation values in hypermultiplets in the 
$(\wedge^2 {\bf r}, {\bf 1})$ representation, the symmetry can be 
broken down to $\SO(32-2r)$ for the case $r=5,8$ [to $\SU(2) \times \SO(28)$ 
for $r=2$]. One can explicitly check that there are $24-2r$ hypermultiplets 
in the SO($32-2r$)-{\bf vect.} representation [$(24-2r)/2$ in the 
$\SU(2) \times \SO(28)$-({\bf 2},{\bf vect.}) representation], in agreement 
with the field-theory calculation. The number of singlet moduli are also 
equal to the sum of the vector bundle moduli and $h^{1,1}=20$ $K3$ moduli.
Since the two singlet hypermultiplets in the untwisted sector are  
genuine $K3$ moduli, remaining 18 are from the 9 twisted sectors. Thus, 
roughly speaking, each twisted sector at $\C^2/\Z_3$ has two hypermultiplets 
for the $K3$ moduli and all the other singlet hypermultiplets in each twisted 
sector correspond to the vector bundle moduli. This is in good agreement 
because two 2-cycles emerge from the blow up of a $\C^2/\Z_3$
singularity (see footnote~\ref{fn:blowup}).
The geometry $T^4/\Z_3$ orbifold is a limit of a $K3$-manifold, where 
$2 \times 9$ 2-cycles are collapsed. The toroidal orbifold 
copmactifications with $V^I_{r=2,5,8}$ in 
(\ref{eq:noWil-k3-r2}--\ref{eq:noWil-k3-r8}) are obtained on top of 
such a singular ``manifold'',  by taking a limit in the 
moduli space of $\SU(2)$, $\SO(10)$ and $\SO(16)$-bundle compactification. 
For the cases with $r=5$ and $8$, this is a limit where the structure 
groups are reduced from $\SO(10)$ ($r=5$) and $\SO(16)$ ($r=8$) 
to $\U(1)$, so that $\SU(5)$ and $\SU(8)$ symmetries are restored. 
Thus, the vector bundles have become line bundles at the orbifold limit. 
The U(1) structure group of the line bundles is also restored 
as a global symmetry there, but we will argue in the 
appendix~\ref{sssec:discuss} that a massless gauge field of the 
$\U(1)$ symmetry does not remain in the spectrum.

{\bf Cases With Discrete Wilson Lines}

Let us now look at toroidal orbifold compactifications with discrete 
Wilson lines $W^I_a \neq 0$. Only a couple of examples are examined 
in the following, and we think that it is enough to see that such 
compactifications are also nothing more than special limits of 
geometric smooth-manifold compactification.

Suppose that an orbifold $T^4/\Z_k$ is a quotient of $\C^2$ by 
a space group generated by a rotation $\sigma$ ($\sigma^k = {\bf id}$) 
and translations $\tau_a$ ($a=1,2,3,4$). Associated with an each element 
of the space group, say, $\tau_a^{m_a} \circ \sigma^n$, is a 
$(\tau_a^{m_a} \circ \sigma^n)$-twisted sector, quantized states of 
worldsheet fields satisfying a boundary condition 
$\Psi(\sigma + 2 \pi) = (\tau_a^{m_a} \circ \sigma^n)(\Psi)(\sigma)$, 
where $\Psi$ denotes worldsheet fields, $Z, \overline{Z}, \psi, 
\overline{\psi}, \lambda$ and $\overline{\lambda}$. 
In the presence of non-trivial discrete Wilson lines, 32 left-moving 
fermions are twisted by a matrix
\begin{equation}
 \gamma_\sigma^n \cdot \beta_a^{m^a} = \diag \left( 
 e^{2\pi \, i (n V^I + m^a W^I_a) } , e^{-2 \pi \, i (n V^I + m^a W^I_a)} \right).
\label{eq:INQ-orbifold}
\end{equation}
A consistency condition corresponding to (\ref{eq:Vafa-orbifold}) 
should be satisfied for each twisted sector, where $V^I$ in 
(\ref{eq:Vafa-orbifold}) is replaced by $n V^I + m^a W^I_a$ mod $\Z$, 
chosen in an interval $[0: 1]$ for 
the $(\tau_a^{m^a} \circ \sigma^n)$-twisted sector.

The generators of the space group satisfy algebraic relations such as 
\begin{equation}
 \tau_a^{m_a} \circ \sigma = \sigma \circ \tau_a^{m'_a}. 
\end{equation}
The twist matrices $\gamma_\sigma$, $\beta_a^{m_a}$ and 
$\beta_a^{m'_a}$ corresponding to the generators 
$\sigma$, $\tau_a^{m_a}$ and $\tau_a^{m'_a}$ should also satisfy 
corresponding relations 
\begin{equation}
 m_a W^I_a \equiv m'_a W^I_a \qquad {\rm mod~} 2\pi \Z.
\end{equation}
$(\tau_a^{m_a} \circ \sigma)$-twisted sector is localized 
at a fixed point $x$ satisfying $\sigma^n x + m^a e_a = x$. 
Because of $+\Lambda$ ambiguity in the fixed points $x$, 
twisted sectors are grouped into 
$\Lambda/(\sigma - {\bf id}) \Lambda$. 
Consistency conditions like (\ref{eq:INQ-orbifold}) have to 
be satisfied for $n V^I + m^a W^I_a$ for each one of 
$\Lambda/(\sigma^n {\rm id}) \Lambda$.  

{\bf Example A}: The following choice of the discrete Wilson line 
is consistent with $T^4/\Z_2$ orbifold with $V^I_{r=2}$ in (\ref{eq:Z2-r2-V}):
\begin{equation}
 W^I_1 = \frac{1}{2} (\overbrace{1,\cdots,1}^4,\overbrace{0,\cdots,0}^{12}), 
 \qquad  W^I_{2,3,4} = 0.
\end{equation}
Eight twisted sectors have a twist vector $V^I$, while eight others 
have $(V+W_1)$; they are given (mod $\Z$) by 
\begin{equation}
 V^I_{r=2} \equiv \frac{1}{2} (\overbrace{1,1}^2,\overbrace{0,0}^2,
                        \overbrace{0,\cdots,0}^{12}),  \qquad 
 (V_{r=2} + W_1)^I \equiv \frac{1}{2} (\overbrace{0,0}^2,\overbrace{1,1}^2,
                        \overbrace{0,\cdots,0}^{12}). 
\end{equation}
The unbroken symmetry is $\SO(4) \times \SO(4) \times \SO(24)$ at the 
orbifold limit,\footnote{The unbroken symmetries at fixed points 
(in twisted sectors) are determined by the twist vectors associated 
with the fixed points. The symmetry group at fixed points with the twist $V^I$
and those with $(V+W_1)^I$ are different subgroups of SO(32), though they 
are both $\SO(4) \times \SO(28)$.} but it can be broken down to SO(24) 
by turning on vev's in some of hypermultiplets. 
Each fixed point has one massless hypermultiplet 
in the SO(24)-{\bf vect.} representation, while such multiplet is absent 
in the untwisted sector. Thus, there are overall 16 hypermultiplets in 
the vector representation, which agrees with the multiplicity $(24 - 2r)$ 
in the case of $r=4$ of the smooth-manifold calculation, with an 
SO(8) bundle and SO(24) unbroken symmetry. The massless spectrum 
calculated through the orbifold technique have 136 SO(24) singlets, 
after the $\SO(4) \times \SO(4)$ symmetry breaking absorbs 12 
hypermultiplets. This agrees with the sum of the number of 
vector bundle moduli, $24 (2r-2) - r(2r-1)=116$, and of the K3 moduli, 20.
Thus, this orbifold compactification can be regarded as a limit of 
smooth K3 manifold compactification with a rank-4 bundle. Even a toroidal 
orbifold with non-trivial discrete Wilson line is regarded as a limit 
of a smooth-manifold compactification with a vector bundle.
Not only the moduli spaces of rank-2, 5, 6, 8 bundles but also that 
of rank-4 bundle contains an orbifold point.

{\bf Example B}: The $T^4/\Z_2$ orbifold with the twist $V^I_{r=2}$ 
in (\ref{eq:Z2-r2-V}) is also consistent with the following discrete 
ilson lines:
\begin{equation}
W^I_1=\frac{1}{2} (\overbrace{1,1}^2,\overbrace{1,1}^2,\overbrace{0,0}^2,
                   \overbrace{0,\cdots,0}^{10}), \quad 
W^I_2=\frac{1}{2} (\overbrace{1,1}^2,\overbrace{0,0}^2,\overbrace{1,1}^2,
                   \overbrace{0,\cdots,0}^{10}), \quad W^I_{3,4}=0,
\label{eq:Z2-r6-W}
\end{equation}
The sixteen fixed points of $T^4/\Z_2$ are classified into 4 groups 
of four fixed points, and each group has its own twist vector given by 
\begin{eqnarray}
 (V+W_2)^I \equiv \frac{1}{2} 
       (\overbrace{0,0}^2,\overbrace{0,0}^2,\overbrace{1,1}^2,
                         \overbrace{0,\cdots,0}^{10}),  &  &
 (V+W_1+W_2)^I \equiv \frac{1}{2} 
       (\overbrace{1,1}^2,\overbrace{1,1}^2,\overbrace{1,1}^2,
                         \overbrace{0,\cdots,0}^{10}), \\
 V^I \equiv \frac{1}{2} 
       (\overbrace{1,1}^2,\overbrace{0,0}^2,\overbrace{0,0}^2,
                         \overbrace{0,\cdots,0}^{10}),  &  &
 (V+W_1)^I \equiv \frac{1}{2} 
       (\overbrace{0,0}^2,\overbrace{1,1}^2,\overbrace{0,0}^2,
                         \overbrace{0,\cdots,0}^{10}).
\end{eqnarray}
The unbroken symmetry is $\SO(4) \times \SO(4) \times \SO(4) \times \SO(20)$
at the orbifold limit, which can be broken down to SO(20) by giving 
vev's in some of hypermultiplets. Massless hypermultiplets in the 
SO(20)-{\bf vect.} 
representation are not found in the untwisted sector or in the four 
twisted sectors with the twist vector $(V+W_1+W_2)^I$. The twelve other 
twisted sectors, whose twist vectors are $V$, $V+W_1$ and $V+W_2$, have
one massless SO(20)-{\bf vect.} hypermultiplet each, and there are twelve 
as a whole. This multiplicity agrees with the smooth-manifold calculation 
of the rank-6 bundle, $[24 - 2r]=12$. The orbifold calculation yields 
194 SO(20)-singlet hypermultiplets (after Higgsing $\SO(4) \times \SO(4) 
\times \SO(4)$), which agrees with the sum of 174 vector bundle moduli 
and 20 K3 moduli of the smooth-manifold calculation.

Thus, the $T^4/\Z_2$ orbifold with $V_{r=6}^I$ and $W^I=0$ and with 
$V_{r=2}^I$ and $W^I$ in (\ref{eq:Z2-r6-W}) are both regarded as special 
limits in the moduli space of $\SO(12)$ vector bundle, limits where 
the structure group is reduced and the unbroken symmetry is enhanced.
In the case with $V^I_{r=6}$ and $W^I=0$, instantons are squeezed 
in the U(1) generated by a charge vector 
${\bf q}= \diag (V^I_{r=6},-V^I_{r=6})$ at all the 16 collapsed
2-cycles. In the case with $V^I_{r=2}$ and the discrete 
Wilson lines in (\ref{eq:Z2-r6-W}), however, they are squeezed 
in a U(1) subgroup generated by 
${\bf q} = \diag ((V_{r=2}+m^a W_a)^I,-(V_{r=2}+m^a W_a)^I )$ 
at the collapsed 2-cycles at $(m^a e_a)/2$; the charge vector ${\bf q}$ 
can be different at different collapsed 2-cycles.

{\bf Example C}: One can introduce discrete Wilson lines in a $T^4/\Z_3$ 
orbifold with the twist $V^I_{r=2}$ as follows:
\begin{equation}
 W^I_1=W^I_2 = \frac{1}{3} (\overbrace{0,0}^2,\overbrace{1,1,1}^3,
            \overbrace{0,\cdots,0}^{11}), \quad W^I_{3,4} = 0.
\label{eq:Z3-r5-W}
\end{equation}
The nine fixed points (twisted sectors) are grouped into 3 sets of 
three fixed points (twisted sectors) whose twist vectors are 
\begin{eqnarray}
\sigma{\rm -twisted}: & V^I_{r=2} & = \frac{1}{3} (1,1,0,0,0,\overbrace{0,\cdots ,0}^{11}), \\
(\tau_1 \cdot \sigma){\rm -twisted}: & (V_{r=2}+W_1)^I & 
      = \frac{1}{3} (1,1,1,1,1,\overbrace{0,\cdots ,0}^{11}), \\
(\tau_1\cdot \tau_2 \cdot \sigma){\rm -twisted}: & 
    (V_{r=2}+W_1+W_2)^I  & = \frac{1}{3} (1,1,2,2,2,\overbrace{0,\cdots ,0}^{11}).
\end{eqnarray}
The unbroken symmetry at the orbifold limit is $\SU(2) \times \SU(3) \times
\SO(22) \left[\times \U(1) \times \U(1) \right]$, but all the factors 
other than SO(22) can be Higgsed away. The total number of massless 
hypermultiplets in the SO(22)-{\bf vect.} representation\footnote{They 
come from two from each fixed point at $m(e_3 + e_4)/3$, 
one from each fixed point at either $(2e_1 + e_2+m(e_3+e_4))/3$ or 
$(e_1+2e_2+m(e_3+e_4))/3$ and 2 from the untwisted sector.} is 14, 
in agreement with the smooth-manifold result for a rank 5 bundle 
(and the unbroken SO(22) symmetry). 
The number of SO(22)-singlet hypermultiplets of this toroidal 
orbifold also agrees with the smooth-manifold calculation. 

The $T^4/\Z_3$ orbifold with $V_{r=5}$ and $W_a = 0$ and 
with $V_{r=2}$ and $W_a$ given in (\ref{eq:Z3-r5-W}) are both special 
points of the moduli space of K3 compactification with a rank-5 vector
bundle. The SO(10) instantons are squeezed in a U(1) subgroup generated 
by ${\bf q} = \diag (V_{r=5},-V_{r=5})$ on all of nine collapsed $\C_2/\Z_3$ 
singularities of a K3 manifold in the case without a discrete Wilson line, 
whereas they are squeezed in 3 different U(1) subgroups at 3 different 
groups of $\C^2/\Z_3$ singularities in the case with the discrete Wilson 
lines (\ref{eq:Z3-r5-W}):
\begin{eqnarray}
\U(1) {\rm ~along~}  {\bf q} = \diag \left(V_{r=2},-V_{r=2} \right) & {\rm at} 
  & \frac{m}{3} (e_3 + e_4), \\
  {\bf q} = \diag \left((V_{r=2}+W_1),-(V_{r=2}+W_1) \right) 
 & {\rm at} & \frac{1}{3} (2 e_1 + e_2 + m(e_3 + e_4)), \\
  {\bf q} = \diag \left((V_{r=2}+W_1+W_2),-(V_{r=2}+W_1+W_2) 
  \right) & {\rm at} & \frac{1}{3} (e_1 + 2 e_2 + m (e_3 + e_4) ). 
\end{eqnarray}
The moduli space of rank-5 bundle compactification contains more 
orbifold points than the two explicitly described above; 
$W^I_1=W^I_2$ can be multiplied by a factor of 2, and $W^I_3=W^I_4$ can 
also be non-zero. At the toroidal orbifold limits with non-trivial 
discrete Wilson lines, the U(1) subgroups in which the instantons are 
squeezed (that is, the structure group of line bundles) 
can be different from one singularity to another. Variety 
of the choice of $W^I_a$ correspond to the variety of finding such 
U(1) subgroups in which the instantons are squeezed. Apart from that, 
there is no essential difference between toroidal orbifolds with or 
without discrete Wilson lines. They are all special limits of 
a simply-connected K3-manifold compactification with a vector
bundle on it.

\subsubsection{Discussion}
\label{sssec:discuss}

We have seen that the toroidal orbifold compactifications of
the Heterotic theory corresponds to special points in the moduli 
space of compactifications with Calabi--Yau and a vector bundle 
on it. At the orbifold points, the structure group of the vector 
bundle is reduced and an unbroken symmetry is enhanced. 
This interpretation holds true regardless of ``discrete Wilson lines'' 
are used or not.
If the twist vectors $V^I$'s and $W^I_a$'s are arranged so that 
U(1) symmetries are left, then the structure group of the bundle 
contains the U(1) symmetries. That is, the bundle contains line bundles 
at such orbifold limits. 

It is important to note that U(1) symmetries in effective theory 
below the Kaluza--Klein scale does not imply that the low-energy
spectrum has corresponding massless vector field. This is why 
we put all the U(1) factors in brackets in the examples of 
toroidal orbifolds in the appendix~\ref{sssec:spec-orbifold}.
If we label multiple U(1) factors of the structure group 
at various fixed points by $a,b$, then the effective lagrangian 
contains 
\begin{equation}
 - \frac{1}{2 g^2} C_{ab} (\partial A^a) (\partial A^b) 
 + \frac{1}{2} G_{kl}(T,T^\dagger) Q^k_a A^a \; Q^l A^b, 
\label{eq:B-A-Het2}
\end{equation}
where the second term comes from (\ref{eq:B-A-Het}).
At the orbifold limits, U(1) symmetries in the directions 
spanned by ${\bf q}$'s may be preserved as global symmetries, 
but the gauge fields acquire mass terms from the second term 
in the effective action above.
The $B$-field fluctuations $b^k \omega_k$ in (\ref{eq:B-A-Het}) 
will be played by twisted sector fields in orbifold language.\footnote{
It is desirable to confirm by explicit orbifold calculations 
that such couplings do exist, but we do not do this in this article.}
Multiple U(1) gauge fields acquire large masses from the 
Stuckelberg form interactions in Heterotic compactifications, 
and toroidal orbifolds with or without ``discrete Wilson lines'' 
are not exceptions.

If orbifold projection conditions are to be used in breaking 
the $\SU(5)_{\rm GUT}$ symmetry down to 
$\SU(3)_C \times \SU(2)_L \times \U(1)_Y$ symmetry in Heterotic theory, 
then such toroidal orbifold compactifications are regarded as limits
where vector bundle has a structure group 
$\U(1) \times \U(1) \times \cdots \subset E_8$. 
Whether the discrete Wilson lines are used or not does not make an 
essential difference in this argument.
The charge vector for the hypercharge ${\bf q}_Y$ is not orthogonal to 
all the charge vectors of the structure groups of the line bundles 
above. Thus, the $\U(1)_Y$ vector field also has a large mass term 
through (\ref{eq:B-A-Het}, \ref{eq:B-A-Het2}) in such toroidal
orbifolds. 

The Stuckelberg coupling with the dilaton chiral multiplet vanishes 
in compactifications of the Heterotic $E_8 \times E'_8$, as long as 
the $\U(1)_Y[\SU(3)_C]^2$ mixed gauge anomaly vanishes. The $\U(1)_Y$ 
symmetry can be preserved (approximately) as a global symmetry in 
low-energy effective theory, if the vev's of vector bundle moduli 
(twisted/untwisted sector fields) are chose appropriately.
Yet, there may not be massless $\U(1)_Y$ gauge field in the low-energy
spectrum, because of the Stuckelberg coupling with the K\"{a}hler 
chiral multiplets (twisted sector fields).

The main text of this article also proposes an idea of how to 
get out of this $\U(1)_Y$ problem.  If one can find a $\U(1)$ 
symmetry in the hidden sector $E'_8$ that is a structure group 
of a line bundle of compactification, and if the first Chern class
of the $\U(1)$ line bundle is the same as that of $\U(1)_Y$, then 
the their linear combination $\U(1)_{\tilde{Y}}$ remains massless.
If the hidden sector is strongly coupled, then the gauge coupling 
constant of $\U(1)_{\tilde{Y}}$  still satisfies the GUT relation 
approximately.

\subsection{Continuous Wilson Lines as Vector Bundle Moduli}
\label{ssec:continuous}

{\bf Example D}: Let us study a following example, to see the claim 
in the title of this subsection. We consider $T^4/\Z_3$ orbifold, 
and take 
\begin{eqnarray}
\gamma_\sigma & = & \tilde{\gamma}_\sigma \oplus \tilde{\gamma}_\sigma^{T-1}, 
\qquad \tilde{\gamma}_\sigma = \left(\begin{array}{ccc}
   & 1 & \\ & & 1 \\ 1 & &  \end{array}\right) \oplus {\bf 1}_{13 \times 13}, 
  \label{eq:beta-diag-baseA} \\
\beta_{a=1} & = & \tilde{\beta}_{a=1} \oplus \tilde{\beta}_{a=1}^{T-1}, 
\qquad \tilde{\beta}_{a=1} = \diag(e^{i\alpha}, e^{i \beta}, e^{-i(\alpha+\beta)}) 
   \oplus {\bf 1}_{13\times 13}, \\
\beta_{a=1} & = & \tilde{\beta}_{a=1} \oplus \tilde{\beta}_{a=1}^{T-1}, 
\qquad \tilde{\beta}_{a=2} = \diag(e^{-i(\alpha+\beta)}, e^{i \alpha}, e^{i\beta}) 
   \oplus {\bf 1}_{13\times 13}. \label{eq:beta-diag-baseC} 
\end{eqnarray}
Those matrices for the orbifold twists are chosen so that they 
satisfy algebraic relations 
\begin{eqnarray} 
  \sigma \circ \tau_{a=1} \circ \sigma^{-1} = \tau_{a=2} & \rightarrow & 
   \gamma_{\sigma}^{-1} \cdot \beta_{a=1} \cdot \gamma_{\sigma} = 
   \beta_{a=2}, \\
  \sigma \circ \tau_{a=2} \circ \sigma^{-1} = (\tau_{a=1} + \tau_{a=2})^{-1}
  & \rightarrow &     \gamma_{\sigma}^{-1} \cdot \beta_{a=2} \cdot \gamma_{\sigma} = 
   (\beta_{a=2} \cdot \beta_{a=1})^{-1}.
\end{eqnarray}
These relations are satisfied for any values of $\alpha, \beta \in \R$, 
and hence this is called the continuous Wilson lines. Certainly the matrix 
$\beta_{a=1,2}$ are the ordinary Wilson lines on torus $T^4$, in the absence 
of orbifold projection by $\Z_3$. 
This is a typical situation where we have a continuous Wilson line.
Although the continuous Wilson lines $(\alpha, \beta)$ are introduced 
only in one of the two complex planes of $T^4$ for simplicity, continuous 
Wilson lines can be introduced for the other complex plane, too. 
Thus, there are four real-scalar degrees of freedom in the continuous 
Wilson lines in this example. 

This example should correspond to a $\SU(3) \subset \SO(6) \subset \SO(32)$
bundle compactification on a K3 manifold, which leaves 
$\U(1) \times \SO(26)$ unbroken symmetry. Therefore, one should have 
\begin{itemize}
 \item D=6 supergravity multiplet and a D=6 tensor multiplet,
 \item $h^{1,1}=20$ hypermultiplets coming from moduli of K3,
 \item D=6 $\SO(26)$ vector multiplet,
 \item 18 hypermultiplets in the vector representation of $\SO(26)$,
 \item 18 $\SO(26)$-singlet hypermultiplets that are charged under 
     the U(1) symmetry, and
 \item 64 completely neutral hypermultiplets coming from vector bundle moduli. 
\end{itemize}

The spectrum can also be calculated using the standard techniques in 
toroidal orbifolds. Each one of the twisted sectors localized at 
9 $\C^2/\Z_3$ singularity contribute to the spectrum of hypermultiplets 
by 2 in the vector representation, 2 in the U(1) charged ones, and 9 
in the U(1) neutral ones. Thus, all of the hypermultiplets in the 4th 
and 5th items in the above list are accounted for in the orbifold 
calculation. The $9 \times 9$ neutral hypermultiplets account for 
$9 \times 2$ of the K3 moduli hypermultiplets and 
$9 \times 7$ of the vector bundle moduli; 
each $\C^2/\Z_3$ singularity has two hypermultiplet worth of 
resolution/deformation degrees of freedom.
Among the twisted-sector spectrum of neutral hypermultiplets, 
two are still missing in the moduli of K3, and 
one in the bundle moduli.

Gravitational part of the untwisted sector gives rise 
to two neutral hypermuliplets, and hence all the 20 hypermultiplets 
for the K3 moduli are recovered from toroidal orbifold calculation. 
The SU(3)-adjoint part of the untwisted sector leaves one massless 
hypermultiplets, and this is identified with the remaining one 
vector bundle moduli. This hypermultiplet takes values in the diagonal 
entries of $3 \times 3$ matrix in the basis that diagonalises 
$\tilde{\beta}$ as in (\ref{eq:beta-diag-baseA}--\ref{eq:beta-diag-baseC}).

One can further see from the orbifold calculation that two 
more hypermultiplets become massless if $\alpha = \beta = 0$, 
and the unbroken symmetry is enhanced 
to $\SO(26) \times \U(1) \times \U(1) \times \U(1)$.
This phenomenon is better understood in a frame that diagonalizes 
the twisting matrix $\tilde{\gamma}_\sigma$ rather than 
$\tilde{\beta}_{a=1,2}$. Generators of $\tilde{\beta}_{a=1,2}$ and 
the hypermultiplets from the untwisted sector take their values 
now in off-diagonal entries of the $3 \times 3$ matrix of adjoint 
$\SU(3)$, and the symmetry breaking $\U(1) \times \U(1) \times \U(1) 
\rightarrow \U(1)$ is understood as the Higgs mechanism due to the 
vev in the untwisted-sector hypermultiplet. Put another way, 
vev's in the untwisted sector hypermultiplet correspond to deformation 
of vector bundle that enlarges the structure group from 
$\U(1) \times \U(1)$ to $\SU(3)$. That is, the continuous Wilson line 
(and the vev's of the untwisted sector hypermultiplets) 
studied in this example corresponds a part of vector bundle moduli 
explained above. 

Continuous Wilson lines exist in cases where the twisting matrix 
$\gamma_\sigma$ acts as permutation. When a basis is chosen so that 
$\gamma_\sigma$ is diagonal, generators of the continuous Wilson lines 
$\beta$ becomes off-diagonal, and the off-diagonal vev's enlarge the 
structure group of vector bundle.

\end{document}